\documentclass[10pt]{article}
\usepackage{fullpage,citesort,epsfig,psfrag,graphics,amsbsy}
\usepackage[normal,small]{caption}
\newcommand{\beq}{\begin{equation}}
\newcommand{\eeq}{\end{equation}}
\newcommand{\bea}{\begin{eqnarray}}
\newcommand{\eea}{\end{eqnarray}}

\newcommand{\Tr}{{\rm Tr}}
\newcommand{\be}{\begin{equation}}
\newcommand{\ee}{\end{equation}}
\newcommand{\bq}{\begin{eqnarray}}
\newcommand{\eq}{\end{eqnarray}}

\newcommand{\ie}{{\it i.e.\ }}
\def\math{\mathsurround=0pt }
\def\leftrightarrowfill{$\math \mathord\leftarrow \mkern-6mu \cleaders\hbox{$\mkern-2mu \mathord- \mkern-2mu$}\hfill
 \mkern-6mu \mathord\rightarrow$}
\def\overleftrightarrow#1{\vbox{\ialign{##\crcr
     \leftrightarrowfill\crcr\noalign{\kern-1pt\nointerlineskip}
     $\hfil\displaystyle{#1}\hfil$\crcr}}}

\let\a=\alpha \let\b=\beta \let\g=\gamma \let\d=\delta 
    
\let\l=\lambda \let\m=\mu   \let\p=\pi 
     
 \let\G=\Gamma \let\D=\Delta  
  \let\S=\Sigma

\def\nn{\nonumber} \def\bd{\begin{document}} \def\ed{\end{document}}
\def\ds{\documentstyle} \let\fr=\frac \let\bl=\bigl \let\br=\bigr
\let\Br=\Bigr \let\Bl=\Bigl
\let\bm=\bibitem
\let\na=\nabla
\let\pa=\partial \let\ov=\overline
\def\ft#1#2{{\textstyle{{\scriptstyle #1}\over {\scriptstyle #2}}}}
\def\fft#1#2{{#1 \over #2}}
\def\vp{\varphi}
\def\sst#1{{\scriptscriptstyle #1}}
\def\oneone{\rlap 1\mkern4mu{\rm l}}
\def\td{\tilde}
\def\wtd{\widetilde}
\def\dalemb#1#2{{\vbox{\hrule height .#2pt
        \hbox{\vrule width.#2pt height#1pt \kern#1pt
                \vrule width.#2pt}
        \hrule height.#2pt}}}
\def\square{\mathord{\dalemb{6.8}{7}\hbox{\hskip1pt}}}
\def\wtd{\widetilde}
\def\R{\rlap{\rm I}\mkern3mu{\rm R}}
\def\im{{\rm i}}
\def\tilg{\tilde{g}}
\def\tilF{\tilde{F}}
\def\tilA{\tilde{A}}
\def\varf{\varphi}
\def\tilf{\tilde{\phi}}
\def\tilh{\tilde{h}}
\def\rme{{\rm e}}
\def\ep{\epsilon}
\def\0{{(0)}}
\def\9{{(9)}}
\def\8{{(8)}}
\def\7{{(7)}}
\def\6{{(6)}}
\def\5{{(5)}}
\def\4{{(4)}}
\def\3{{(3)}}
\def\2{{(2)}}
\def\1{{(1)}}
\newcommand{\trace}{{\rm Tr}}
\newcommand{\ub}{\overline{U}}
\newcommand{\vb}{\overline{V}}
\newcommand{\uh}{\widehat{U}}
\newcommand{\vh}{\widehat{V}}
\newcommand{\ubh}{\overline{\widehat{U}}}
\newcommand{\vbh}{\overline{\widehat{V}}}
\newcommand{\lb}{\bar{\l}}
\newcommand{\Fb}{\overline{F}}
\newcommand{\Fh}{\widehat{F}}
\newcommand{\Fbh}{\overline{\widehat{F}}}
\newcommand{\Ab}{\overline{A}}
\newcommand{\Ah}{\widehat{A}}
\newcommand{\Abh}{\overline{\widehat{A}}}
\newcommand{\Gb}{\overline{G}}
\newcommand{\Gh}{\widehat{G}}
\newcommand{\Gbh}{\overline{\widehat{G}}}
\newcommand{\Pb}{\overline{P}}
\newcommand{\Ph}{\widehat{P}}
\newcommand{\Pbh}{\overline{\widehat{P}}}
\newcommand{\Qb}{\overline{Q}}
\newcommand{\Qh}{\widehat{Q}}
\newcommand{\Qbh}{\overline{\widehat{Q}}}
\newcommand{\Bb}{\overline{B}}
\newcommand{\Bh}{\widehat{B}}
\newcommand{\Bbh}{\overline{\widehat{B}}}
\newcommand{\fhns}{\hat{F}^{\rm (NS)}}
\newcommand{\fhrr}{\hat{F}^{\rm (RR)}}
\newcommand{\ahns}{\hat{A}^{\rm (NS)}}
\newcommand{\ahrr}{\hat{A}^{\rm (RR)}}
\newcommand{\hhrr}{\hat{H}^{\rm (RR)}}
\newcommand{\hchi}{\hat{\chi}}
\newcommand{\hphi}{\hat{\phi}}
\newcommand{\htau}{\hat{\tau}}
\newcommand{\cG}{{\cal G}}
\newcommand{\cGb}{\overline{{\cal G}}}
\newcommand{\cH}{{\cal H}}
\newcommand{\cP}{{\cal P}}
\newcommand{\cPb}{\overline{{\cal P}}}
\newcommand{\cQ}{{\cal Q}}
\newcommand{\cQb}{\overline{{\cal Q}}}
\newcommand{\cM}{{\cal M}}
\newcommand{\cN}{{\cal N}}
\newcommand{\cO}{{\cal O}}
\newcommand{\cD}{{\cal D}}
\newcommand{\cL}{{\cal L}}

\begin{document}
\setlength{\captionmargin}{20pt}

\renewcommand{\thefootnote}{\fnsymbol{footnote}}
\begin{titlepage}
\begin{flushright}
UFIFT-HEP-29-02\\
hep-th/0209102
\end{flushright}

\vskip 3cm

\begin{center}
\begin{Large}
{\bf BT Worldsheet for
Supersymmetric  Gauge Theories\footnote{
This work was supported in part by the Department
of Energy under Grant No. DE-FG02-97ER-41029. 
}}
\end{Large}

\vskip 2cm
{\large Skuli Gudmundsson\footnote{E-mail  address: 
{\tt skulig@phys.ufl.edu}}, 
 Charles B. Thorn\footnote{E-mail  address: {\tt thorn@phys.ufl.edu}},
and Tuan A. Tran\footnote{E-mail  address: {\tt tuan@phys.ufl.edu}}}
\vskip0.20cm
{\it Institute for Fundamental Theory\\
Department of Physics, University of Florida,
Gainesville FL 32611}

(\today)

\vskip 1.0cm
\end{center}

\begin{abstract}\noindent
We extend the Bardakci-Thorn (BT) worldsheet formalism to 
supersymmetric non-abelian gauge theory. Our method covers
the cases of ${\cal N}=1,2,4$ extended supersymmetry.
This task requires the
introduction of spinor valued Grassmann variables on the worldsheet
analogous to those of the supersymmetric formulation
of superstring theory. As in the pure Yang-Mills case,
the worldsheet formalism automatically generates the correct
quartic vertices from the cubic vertices.
We also discuss coupling renormalization to one loop order.
\end{abstract}
\vfill
\end{titlepage}
\section{Introduction}
\label{sec1}
Last year Bardakci and one of us \cite{bardakcit} proposed a method
for mapping each individual planar Feynman diagram of
the large $N_c$ limit \cite{thooftlargen}
of a matrix quantum field
theory onto the evolution amplitude of a ``topological''
worldsheet dynamical system defined on a light-cone worldsheet
\cite{goddardgrt,mandelstam,gilest}. 
The sum over all planar diagrams is then accomplished through the
introduction of an Ising-like spin system on this same worldsheet,
which is coupled to the target space worldsheet fields.
This interacting system can be thought of as noninteracting
string propagating on a highly non-trivial background represented
by the Ising-like spins. The initial proposal was developed
for $\Tr\phi^3$ scalar field theory, but the formalism was soon 
extended to the case of pure Yang-Mills theory \cite{thornsheet}.

Extracting the physics of the large $N_c$ limit of pure 
Yang-Mills theory is probably the most exciting potential
application of this new formalism. It would be the zeroth
order of a systematic expansion of QCD in powers of $1/N_c$, 
which would provide theoretical physics with
an analytic understanding of the spectrum and structure of glueballs
and, with the inclusion of quarks, that of other hadrons.
A mean field method for capturing the nonperturbative
physics of the worldsheet formalism has already been
initiated \cite{bardakcitmean}. If the idea of
a quasi-perturbative gluon chain model of the 
quark confining flux tube
\cite{greensitet,thooftpconfine} is indeed viable, the BT
worldsheet formalism should be the ideal setting for its
development. This is because the stringy features of the
theory are extracted directly from the pertubative diagrams. 

However, in this article, we are interested in testing the 
BT formalism by extending it to theories for which 
a stringy description has been understood from other
points of view. In particular, Maldacena \cite{maldacena} has proposed
that the large $N_c$
limit of ${\cal N}=4$ supersymmetric Yang-Mills theory is
equivalent to a noninteracting
string theory on an AdS$_5\times$S$_5$ background
\cite{maldacena,gubserkp,wittenholog}. Unfortunately,
calculations in this approach have generally been tractable
only at large 't Hooft coupling $N_cg_s^2\to\infty$. Since the
BT worldsheet is based on {\it weak} 't Hooft coupling, it
should provide complementary insight into the workings
of Maldacena duality.
Therefore, in this article we extend the formalism to include
fermions and, in particular, our method covers the cases of supersymmetric
gauge theories with ${\cal N}=1,2,4$. Study
of the ${\cal N}=4$ case should then throw new light on
the Maldacena conjecture.
We note in passing two earlier works that share similar goals to
ours but differ in method. The first \cite{nastases} is an effort to abstract
a covariant worldsheet formalism from the planar graphs
of ${\cal N}=4$ supersymmetric gauge theories.
A more recent work on the pp-wave limit of AdS$_5\times$S$_5$
has led to another intriguing interpolation between the strong
and weak coupling regimes \cite{berensteinmn}.  

The worldsheet construction of Ref~\cite{bardakcit}
exploits light-cone coordinates\footnote{ 
The light-cone components of a Minkowski vector $v^\mu$ are defined
as $v^\pm=(v^0\pm v^{D-1})/\sqrt2$, with the remaining (transverse) components
of $v^\mu$ distinguished by Latin indices, or
as a vector by bold-face type. The Lorentz invariant
scalar product of two four vectors $v,w$ is written
$v\cdot w=\boldsymbol{ v}\cdot\boldsymbol{ w}-v^+w^--v^-w^+$.}. On the light-cone
$x^+$ is the quantum evolution parameter, and 
the Hamiltonian conjugate to this time is
$p^-$. A massless on-shell particle thus has the ``energy''
$p^-=\boldsymbol{ p}^2/2p^+$.
The construction begins with the identification of a
worldsheet system that reproduces 
the mixed representation ($x^+,p^+,\boldsymbol{ p}$) of the propagator 
of a free massless scalar field \cite{thooftlargen}:
\begin{eqnarray}
\exp\left\{-\tau
{\boldsymbol{ p}^2\over2p^+}\right\}&=&
\int DcDbD\boldsymbol{ q}\ 
\exp\left\{-\int_0^T d\tau \int_0^{p^+}d\sigma\ \left[{\boldsymbol{ q}^{\prime2}\over2}
-b^\prime c^\prime\right]\right\}.
\label{freeworldsheet}
\end{eqnarray}
Here the prime denotes $\partial/\partial\sigma$, and we are working
with imaginary $x^+$ or real $\tau\equiv ix^+$.  
With lightcone parametrization the worldsheet is just a
rectangle of width $p^+$ and length $T$.
In the path integral the worldsheet fields include
the target space field $\boldsymbol{ q}(\sigma,\tau)$, 
with Dirichlet boundary conditions
constrained by 
$ \boldsymbol{q}(p^+,\tau)-\boldsymbol{ q}(0,\tau)=\boldsymbol{ p}$
the total transverse momentum of the system. 
The derivative of $\boldsymbol{ q}$  is the density of transverse momentum
on a bit of worldsheet: that is $\boldsymbol{ q}^\prime d\sigma$
is the transverse momentum carried by the element $d\sigma$.
The anticommuting ghost fields $b,c$  ensure that the 
correct measure is obtained. 

The light-cone form of any field theoretic propagator, whether
it is for a scalar, fermion, or gauge field is always
simply the scalar propagator times a Kronecker delta that
describes the flow of spin and other internal quantum numbers.
Thus the expression (\ref{freeworldsheet}) is a universal
part of the worldsheet construction for any field. When internal
degrees of freedom are also present, however, one must also
give a local worldsheet description of them. In the case of
pure Yang-Mills, this was accomplished by introducing
a transverse vector valued Grassmann odd worldsheet
field $S^k(\sigma,\tau)$ \cite{thornsheet}.
The absence of bulk dynamical variables
on the worldsheet is evident from the absence of $\dot{\boldsymbol{q}}$
dependence in the action. This means that the bulk fields
are determined by their boundary values, which is the sense in which
we describe the worldsheet system as topological. However boundary
dynamics is implicit in the Dirichlet boundary conditions which
correlate different time slices.  

Note that
a factor of $1/2p^+$ present in the usual bosonic propagator has
been removed: it must therefore be included in the
definition of the vertices. By convention we introduce
$m$, a unit of $p^+$, and include the dimensionless
factor $m/p^+$ in the {\it earlier} of the two vertices connected by the
propagator, and a factor $1/\sqrt{2m}$ in {\it each} of
the two vertices. A cubic vertex is represented on the rectangular
worldsheet just described by the appearance (or disappearance)
at some time of an interior Dirichlet boundary at fixed $\sigma$. The
value of $\boldsymbol{ q}$ on this boundary governs how the transverse momentum
is shared among the particles. For example, a fission vertex
is the appearance of a solid line, say at $\sigma=p^+_1$, 
representing the new boundary.
Before this occurs the system is a single particle with momentum
$\boldsymbol{ p}=\boldsymbol{ q}(p^+)-\boldsymbol{ q}(0)$. Afterwards the system is
two particles with momenta $\boldsymbol{ p}_1=\boldsymbol{ q}(p_1^+)-\boldsymbol{ q}(0)$,
$\boldsymbol{ p}_2=\boldsymbol{ q}(p^+)-\boldsymbol{ q}(p^+_1)$. If the new boundary
line subsequently terminates, the diagram contains an extra loop.
Thus the sum over all planar diagrams in a theory with only cubic
vertices is just the sum over all ways of inserting such boundary
lines within the worldsheet. This sum can be accomplished technically 
by discretizing $\sigma=lm$ and $\tau=ka$ as in \cite{gilest} and
introducing an Ising spin variable on each temporal bond that
keeps track of whether it is part of an interior boundary (drawn
as a solid line) or not (drawn as a dotted line).
The technical details of this procedure are described in 
\cite{bardakcit,thornsheet}.

Quartic and higher point vertices would seem to spoil this 
nice worldsheet picture by introducing nonlocal
features into the worldsheet description. It is therefore very
satisfying that the quartic interactions required in
Yang-Mills theory are automatically generated by the worldsheet formalism
from the presence of two cubic vertices, which are
linear in the transverse momenta \cite{thornsheet}.
Note that this does {\it not} happen in purely scalar field theory 
where quartic vertices would require
a nonlocal worldsheet dynamics. The status of the worldsheet
description of fermion fields also needs to be evaluated. 
Since supersymmetry 
requires the presence of fermion fields and extended
supersymmetry the presence of additional scalar fields, 
we face the important
question: which supersymmetric theories can
be given a local worldsheet description? This article is
devoted to answering this question.
 
A concise and very convenient way to specify the field content
and couplings of a gauge theory with extended supersymmetry,
is to begin with a ${\cal N}=1$ gauge theory in higher dimensions $D>4$
and then apply dimensional reduction. This means that all the
fields are required to be independent of the $D-4$ extra
coordinates. Then the extra components of the gauge field
become scalar fields from the four dimensional point of view,
and the higher dimensional representation of the Dirac
matrices account for the multiplicity of spin 1/2 fields needed
for the extended supersymmetry. In this way, the ${\cal N}=2$
supersymmetric gauge theory descends from ${\cal N}=1$
in $D=6$ dimensions and the ${\cal N}=4$ case descends from
${\cal N}=1$ in $D=10$ dimensions. Applying this method to the
worldsheet construction, the first step is to promote the
worldsheet field $\boldsymbol{ q}$ to a $D-2$ component vector. One must
at the same time supplement the ghost system with a new $b,c$
pair for each pair of new dimensions. Once this
is done dimensional reduction is simply the imposition of true
Dirichlet boundary conditions on the extra components of $\boldsymbol{ q}$:
$q^k=0$ on {\it all} worldsheet boundaries for $k=3,\ldots,D-2$.
In other words, in the language of
string theory we restrict the fields to a three brane.
Of course, in addition to the new components of $\boldsymbol{ q}$,
one must also add new components to the Grassmann spin
variables that are monitoring the flow of internal degrees
of freedom through the worldsheet.

One might at first think that the new components of $\boldsymbol{ q}$ 
are complete dummies contributing nothing new to the
dynamics, leaving only the extra Grassmann
variables to enrich the physics of the system. 
After all, by construction the bulk variables
have no dynamical significance, and by setting the boundary
values of these extra components to zero, it seems one has
completely eliminated their dynamical relevance. However, this
is not the case because the {\it fluctuations} of the
$\boldsymbol{ q}$ variables are instrumental in generating the
quartic vertices from pairs of cubic vertices. Since some of
the new quartic interactions exchange the $O(D-4)$
quantum numbers carried by the scalars, it
is clear that a local worldsheet description will {\it
require} that these extra components of $\boldsymbol{ q}$ be present. 

We begin our work in the next section by using the
light-cone Feynman rules for gauge fields in general dimension
$D\geq4$ to construct the worldsheet system that will reproduce
all the planar diagrams in four dimensions containing the
gauge particles and scalars and their cubic vertices
necessary for ${\cal N}=1,2,4$ gauge theories. In section 3
we do the same for the fermion fields and their cubic
interactions with the gauge particles and scalars.
In Section 4 we turn to the quartic vertices.
We show that the basic mechanism for their generation,
discovered in \cite{thornsheet}, applies here as well.
However, we also find an interesting limitation to its
applicability. The coefficients of some of the 
generated quartics are dimension dependent, whereas the
desired ones are not. One can arrange the correct
values of these coefficients only if one dimensionally
reduces to $4$ or less space-time dimensions. 

In Section 5, we present a system of Grassmann 
worldsheet fields that locally describes the 
flow of internal degrees of freedom through planar
diagrams. We find that it is sufficient to introduce
two sets of spinor-valued variables $S^a,{\bar S}^{b}$,
where $a,b$ are the spinor indices associated
with the transverse rotation group $O(D-2)$. Vertex insertions
involve either $S^a$ (${\bar S}^b$) for a fermion (anti-fermion) 
entering the vertex or the bilinear 
$S^{b}\gamma^k_{ba}{\bar S}^a$ for a scalar or gauge particle
entering the vertex. Since the number of fermions entering the
vertex is always even, the overall vertex insertion will
be Grassmann even.
Finally in section 6 we present one loop calculations
in enough detail to learn how coupling constant renormalization
works in the worldsheet language. We note several
intriguing features. First it is recalled that
the cancellation of entangled ultraviolet and infrared
divergences familiar in light-cone calculations happens
locally on the world-sheet. Once this cancellation has been
taken into account, the remaining coupling renormalization
also has an interesting local worldsheet interpretation.
In particular, it is found that the cancellations
typical of supersymmetry happen locally. 
Some further discussion and concluding remarks are given in Section 7. 

\section{Gauge Theory in $D$ Dimensions Reduced to 4}
\label{sec2}
As described in the introduction, we will be studying theories
that can be obtained by dimensional reduction from an ${\cal N}=1$
gauge theory in $D$ dimensions. The Lagrangian for such a theory is
just
\bea
{\cal L}
&=&-{1\over4}\Tr F_{\mu\nu}F^{\mu\nu}
+i\Tr\psi^\dagger\alpha^\mu(\partial_\mu\psi-ig\left[A_\mu,\psi\right])\\
F_{\mu\nu}&\equiv&\partial_\mu A_\nu-\partial_\nu A_\mu
-ig\left[A_\mu,A_\nu\right],
\label{d-qcd}
\eea
where $\alpha^\mu\equiv \Gamma^0\Gamma^\mu$ with $\Gamma^\mu$
the $D$ dimensional Dirac gamma matrices.
In this section we concentrate only on the bosonic fields. Fermions will
be discussed in the next section.

We work in light-cone gauge $A_-=0$. After eliminating $A_+$
using the Gauss' law constraint we arrive at the density of
$P^-$
\bea
{\cal P}^-&\equiv&{1\over2}\Tr\partial_iA_j\partial_iA_j-{g^2\over2}\ 
\Tr\Biggl[\left({1\over\partial_-}[A_k,\partial_-A_k]\right)^2
+A_iA_j[A_i,A_j]\Biggr]\nonumber\\
&&\qquad +\ ig \Tr\partial_-A_k\Biggl[A_j{\overleftrightarrow 
{\partial_k\over\partial_-}}A_j
-A_k{\overleftrightarrow {\partial_i\over\partial_-}}A_i
-A_i{\overleftrightarrow{\partial_i\over\partial_-}}A_k
\Biggr],
\eea
so that $H=P^-=\int d\boldsymbol{ x}dx^-{\cal P}^-$.
Here we have introduced the shorthand notation
\bea
X{\overleftrightarrow{\partial_i\over\partial_-}}Y
\equiv X{\partial_i\over\partial_-}Y
-\left({\partial_i\over\partial_-}X\right) Y.
\eea
We easily see that the free propagator is just the scalar
propagator times $\delta_{ij}$. To construct the
worldsheet system the cubic interactions are all-important.

\subsection{Cubic Yang-Mills Vertices in General Dimension}
\label{sec2.1}
We first express the cubic term  $P_1^-$ of $P^-$ in momentum modes:
\bea
A_k=\int {d^{D-1}p\over(2\pi)^{(D-1)/2}\sqrt{2p^+}}(a_k(p)e^{ix\cdot p}
+a^\dagger_k(p)e^{-ix\cdot p}),
\eea
where $d^{D-1}p=dp^+d^{D-2}p\theta(p^+)$ and $x\cdot p=\boldsymbol{ x}\cdot\boldsymbol{ p}
-x^-p^+$. Then we find
\bea
P_1^-&=&-\int {dp_1dp_2dp_3\over\sqrt{|p_1^+p_2^+p_3^+|}}
\left[\Tr\ a_{n_2}^\dagger(-p_2)a_{n_1}^\dagger(-p_1)
a_{n_3}^{\phantom{\dagger}}(p_3) + \Tr\ a_{n_3}^\dagger(-p_3)
a_{n_2}^{\phantom{\dagger}}(p_2)a_{n_1}^{\phantom{\dagger}}(p_1)\right]
\nonumber\\
&&\hskip1cm \times V^{n_1n_2n_3}\delta(p_1+p_2+p_3),
\label{pminus1}
\eea
where $V^{n_1n_2n_3}$ is given by\footnote{
The factor $g/8\pi^{3/2}$ is appropriate for dimensional
reduction to $4$ dimensions. Before reduction it started as 
$g_D2^{-3/2}(2\pi)^{-(D-1)/2}$. To carry out the reduction
one first compactifies each extra dimension so that 
$p\to 2\pi n/L$ and then takes $L\to0$, so only the mode $n=0$
is kept. Then the measure and $a$'s in (\ref{pminus1})
together provide a factor $(2\pi/L)^{(D-4)/2}$, producing
$g_DL^{-(D-4)/2}\pi^{-3/2}/8$. Recalling that the appropriate
coupling in 4 dimensions as $L\to0$ is $g\equiv g_DL^{-(D-4)/2}$,
we arrive at the quoted result.}
\bea
V^{n_1n_2n_3}&=&{g\over8\pi^{3/2}}
\left\{\delta_{n_1n_2}\left(p^+_3\left[{p_1\over p^+_1}
-{p_2\over p^+_2}\right]^{n_3}+p^+_2\left[{p_1\over p^+_1}
-{p_3\over p^+_3}\right]^{n_3}+p^+_1\left[{p_3\over p^+_3}
-{p_2\over p^+_2}\right]^{n_3}\right)\right.\nonumber\\ &&\left.
\hskip.5cm+\delta_{n_1n_3}\left(p^+_2\left[{p_3\over p^+_3}
-{p_1\over p^+_1}\right]^{n_2}+p^+_3\left[{p_2\over p^+_2}
-{p_1\over p^+_1}\right]^{n_2}+p^+_1\left[{p_3\over p^+_3}
-{p_2\over p^+_2}\right]^{n_2}\right)\right.\nonumber\\ &&\left.
\hskip.5cm+\delta_{n_2n_3}\left(p^+_1\left[{p_2\over p^+_2}
-{p_3\over p^+_3}\right]^{n_1}+p^+_3\left[{p_2\over p^+_2}
-{p_1\over p^+_1}\right]^{n_1}+p^+_2\left[{p_1\over p^+_1}
-{p_3\over p^+_3}\right]^{n_1}\right)\right\}.
\label{cartcubic}
\eea
In the expression (\ref{pminus1})
it is understood that the $p^+$ argument of $a$ is always positive,
so it is implied that the range of integration is 
$-p_1^+,-p_2^+>0$.
For practical calculations remember that, when spatial ($p^k,p^+$) 
momentum conservation is taken
into account, all of the momentum differences 
appearing in (\ref{cartcubic}) are proportional to the
single momentum
\bea
\boldsymbol{ K}\equiv p^+_2\boldsymbol
{ p}_1-p^+_1\boldsymbol{ p}_2=p^+_3\boldsymbol{ p}_2-p^+_2\boldsymbol{ p}_3
=p^+_1\boldsymbol{ p}_3-p^+_3\boldsymbol{ p}_1,
\label{defK}
\eea
so we can obtain the dramatic simplification
\bea
V^{n_1n_2n_3}
={g\over4\pi^{3/2}}\left(\delta_{n_1n_2}{K^{n_3}\over p^+_3}
+\delta_{n_1n_3}{K^{n_2}\over p^+_2}
+\delta_{n_2n_3}{K^{n_1}\over p^+_1}\right).
\label{cartcubicsimple}
\eea
However in translating to the BT worldsheet formalism, it
is important to choose in each term a version of $K$ that
allows cancellation of the $1/p^+_r$ factor. We 
shall stick with the original form (\ref{cartcubic})
which makes these choices in a cyclically symmetric manner 
and does not exploit momentum conservation.

To present supersymmetric gauge theory for ${\cal N}>1$, we will, 
in addition to fermions, need additional scalars. These are most simply 
obtained by dimensionally reducing from an ${\cal N}=1$ SUSY
gauge theory in higher dimensions. Let $n=1,2,\ldots,D-2$ label the components
of transverse space. Then
dimensional reduction to 4 dimensions is achieved by setting
$p_k^n=0$ for $n=3,4,\ldots, D-2$, giving $D-4$ scalars, $\phi_k=A^{2+k}$. 
This is, of course,  done not only for
external momenta but also internally for all loop momenta, only the
first two components of which are integrated. When the cubic vertex involves
some of these scalars, only the two scalar-vector
vertex is non-vanishing. Taking the scalars to be particles $1,2$,
so that $n_1,n_2>2$ and $n_3=1,2$, we see that the 
scalar-scalar-vertex only gets contributions from the first line
of Eq.~(\ref{cartcubic}).

Now we consider how to set up the worldsheet system corresponding to
this dimensionally reduced gauge theory. Worldsheet fields 
${q}^n(\sigma,\tau)$ will be introduced for all $D-2$ components.
However, the boundary conditions will vary depending on $n$.
For $n=1,2$ we impose the usual Dirichlet conditions so that
$q^n(p^+)-q^n(0)=p^n$. On the other hand we shall require
the components $n=3,\ldots,D-2$ to strictly vanish on all
boundaries. This is the worldsheet version of dimensional
reduction: the physics resides on a D3-brane in the $q$ space.
As usual there will also be $(D-2)/2$ pairs of ghosts $b_a,c_a$,
to ensure the correct measure.

\begin{figure}[ht]
\begin{center}
\psfrag{'A'}{$A$}
\psfrag{'B'}{$B$}
\psfrag{'C'}{$C$}
\psfrag{'D'}{$D$}
\psfrag{'1'}{$1$}
\psfrag{'2'}{$2$}
\psfrag{'3'}{$3$}
\psfrag{'k'}{$k$}
\psfrag{'l'}{$l$}
\includegraphics[width=10cm]{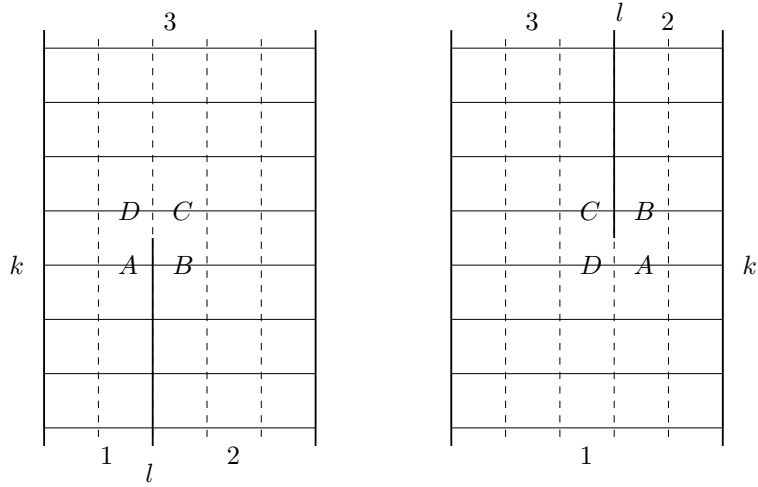}
\end{center}
\caption{Discretized worldsheet for cubic 
fusion and fission vertices showing the possible locations of
momentum insertions $\Delta\boldsymbol{ q}$. We have labeled
the four links surrounding the vertex $A$, $B$, $C$,
$D$. For example
an insertion at link $A$ produces the factor $\boldsymbol{ p}_1/M_1$.
Similarly, insertions at $B$, $C$ produce the analogous factors for
particles 2 and 3 respectively. }
\label{pinserts}
\end{figure}

The worldsheet for a cubic vertex will contain one interior boundary
that terminates within the diagram and extends either to early
or late times (see Fig.\ref{pinserts}). 
As in \cite{thornsheet} the
momentum factors of the cubic vertex are produced by the insertion 
of $\partial q/\partial\sigma$ at suitable points in the
vicinity of the end of the interior boundary. In practice
we put the worldsheet on a grid \cite{gilest} by discretizing $\tau=ka$,
$\sigma=lm$ for $k,l=1,2,\ldots$, $q(\sigma,\tau)\to q_l^k$. 
Since $p^+$ is now restricted
to discrete values, the $p^+$ conserving delta function is
replaced by $1/m$ times a Kronecker delta. Further
the vertex which is a matrix element of $-idx^+P^-_1\to -aP_1^-$.
Thus the vertex function will be
\bea
\Gamma^{n_1n_2n_3}={a\over m}V^{n_1n_2n_3},
\eea
where each $p_k^+\equiv M_km$ is now discrete. A single discretized
insertion $\Delta\boldsymbol{ q}_L=\boldsymbol{ q}^k_{l+1}-\boldsymbol{ q}^k_{l}$, where the link
$L$ is bounded by $l,l+1$, placed anywhere
on the worldsheet of a single string has the expectation value
$\boldsymbol{ p}/M$. Note that since $p^k=0$ for $k=3,\ldots,D-2$, the
cubic vertex will be correctly produced whether 
the index $k$ of $q$ runs over just $k=1,2$ or over the
whole set $k=1,2,\ldots,D-2$. We shall require both choices,
which we shall distinguish with a hat over the $q$ if
$k=1,2$ only, and no hat if it
runs over just the whole range $k=1,\ldots,D-2$.
The complete vertex insertion is then a
sum over three kinds of terms, each with the insertion on
a different string, \ie\ on a different bond $A$, $B$, or $C$
in Fig.~\ref{pinserts}. The factor multiplying each
such insertion involves Kronecker delta's in the polarizations
and is linear in the $M_i$, which we shall write
in terms of those of the two particle state. Thus at a fusion vertex 
we write the total vertex function as
\bea
V_{fusion}^{n_1n_2n_3}&\to&{\bar V}^{n_1n_2n_3}={g\over8\pi^{3/2}}
\left(\delta_{n_1n_2}\left[-M_1{\Delta {q}}_A^{n_3}
+M_2{\Delta {q}}_B^{n_3}
+(M_1-M_2){\Delta {\hat q}}_C^{n_3}\right]\right.\nonumber\\
&&\qquad+\delta_{n_1n_3}\left[M_1{\Delta {q}}_A^{n_2}
-(2M_1+M_2){\Delta {\hat q}}_B^{n_2}
+(M_1+M_2){\Delta {q}}_C^{n_2}\right]\nonumber\\
&&\qquad\left.+\delta_{n_2n_3}\left[(2M_2+M_1){\Delta {\hat q}}_A^{n_1}
-M_2{\Delta {q}}_B^{n_1} 
-(M_1+M_2){\Delta {q}}_C^{n_1}\right]\right).
\label{fusioninsert}
\eea
On the other hand we write the insertion at a fission vertex as
\bea
V_{fission}^{n_1n_2n_3}&\to&V^{n_1n_2n_3}={g\over8\pi^{3/2}}
\left(\delta_{n_1n_2}\left[(M_2+M_3){\Delta {q}}_A^{n_3}
+M_2{\Delta {q}}_B^{n_3}
-(M_3+2M_2){\Delta {\hat q}}_C^{n_3}\right]\right.\nonumber\\
&&\qquad+\delta_{n_1n_3}\left[-(M_2+M_3){\Delta {q}}_A^{n_2}
+(2M_3+M_2){\Delta {\hat q}}_B^{n_2}
-M_3{\Delta {q}}_C^{n_2}\right]\nonumber\\
&&\qquad\left.+\delta_{n_2n_3}\left[(M_2-M_3){\Delta {\hat q}}_A^{n_1}
-M_2{\Delta {q}}_B^{n_1} 
+M_3{\Delta {q}}_C^{n_1}\right]\right).
\label{fissioninsert}
\eea
Remember that in the fission case $M_2$ and $M_3$ are both negative.
Note the occurrence of both hatted and unhatted $q$'s in these
expressions: we shall see that this is essential in order for the fluctuations
of $q$ to produce the correct quartic vertices.
The task of representing the
remaining $M_i$ dependence locally on the 
worldsheet will be handled in the next subsection.
\subsection{Polynomials in $p^+$}
\label{sec2.2}
As in \cite{thornsheet} we handle the factors of $M_i$ of 
Eqs~(\ref{fusioninsert}),~(\ref{fissioninsert})
by more ghost degrees of freedom, $\beta,\gamma$ and 
${\bar\beta},{\bar\gamma}$. 
On each time slice we insert
\begin{eqnarray}
&&\int \prod_{i=1}^{M-1} {d\gamma_id\beta_i}
{d{\bar\gamma}_id{\bar\beta}_i} 
\exp\left\{
\beta_1\gamma_1+\sum_{i=1}^{M-2}
(\beta_{i+1}-\beta_i)(\gamma_{i+1}-\gamma_i)\right.\nonumber\\
&&\hskip4cm\left.+
{\bar\beta}_{M-1}{\bar\gamma}_{M-1}
+\sum_{i=1}^{M-2}({\bar\beta}_{i+1}-{\bar\beta}_i)
({\bar\gamma}_{i+1}-{\bar\gamma}_i)\right\}
=1.
\label{dummyunity}
\end{eqnarray}
Here we have dispensed with the factors of $2\pi$ and $a/m$
present in the $b,c$ path integral.
This insertion is completely harmless because it does nothing. 
But with these dummy-ghost systems 
available, we can locally produce factors of $M_i$
at will as they are needed. 
For example, either $e^{\beta_{M-1}\gamma_{M-1}}$ applied
on the right of a strip of $M$ bits 
or $e^{{\bar\beta}_1{\bar\gamma}_1}$ applied
on the left of the strip produces a factor of $M$. 
At a fusion vertex, the end of
a solid line marks where two strips, 1 to the left of 2, 
join a single larger strip
3. Then an insertion of the first type on strip 1 produces $M_1$, of the
second type on strip 2 produces $M_2$, and the sum of the two  
insertions produces $M_1+M_2=-M_3$. Similarly at a fission
vertex a larger strip 1 joins two strips, 2 to the right
of 3. Then an insertion of the first type on strip 3 produces $-M_3$, 
of the second type on strip 2 produces $-M_2$, and the sum of the two  
insertions produces $-M_2-M_3=M_1$. Thus 
we make the substitutions
\bea
M_1\to e^{\beta_A\gamma_A},\qquad M_2\to e^{{\bar\beta}_B{\bar\gamma}_B},
\qquad{\rm for~fusion~(Eq.~\ref{fusioninsert})},\nonumber\\
M_2\to -e^{\beta_B\gamma_B},\qquad M_3\to -e^{{\bar\beta}_C{\bar\gamma}_C},
\qquad{\rm for~fission~(Eq.~\ref{fissioninsert})}.
\eea
 All together we therefore have 
\bea
{{\bar{V}}^{n_1n_2n_3}}&=&{g\over8\pi^{3/2}}
\left(\delta_{n_1n_2}\left[-e^{\beta_A\gamma_A}{\Delta {q}}_A^{n_3}
+e^{{\bar\beta}_B{\bar\gamma}_B}{\Delta {q}}_B^{n_3}
+(e^{\beta_A\gamma_A}-e^{{\bar\beta}_B{\bar\gamma}_B}){\Delta {\hat q}}_C^{n_3}\right]\right.\nonumber\\
&&\qquad+\delta_{n_1n_3}\left[e^{\beta_A\gamma_A}{\Delta {q}}_A^{n_2}
-(2e^{\beta_A\gamma_A}+e^{{\bar\beta}_B{\bar\gamma}_B})
{\Delta {\hat q}}_B^{n_2}
+(e^{\beta_A\gamma_A}+e^{{\bar\beta}_B{\bar\gamma}_B})
{\Delta {q}}_C^{n_2}\right]\nonumber\\
&&\qquad\left.+\delta_{n_2n_3}\left[(2e^{{\bar\beta}_B{\bar\gamma}_B}
+e^{\beta_A\gamma_A}){\Delta {\hat q}}_A^{n_1}
-e^{{\bar\beta}_B{\bar\gamma}_B}{\Delta {q}}_B^{n_1} 
-(e^{\beta_A\gamma_A}+e^{{\bar\beta}_B{\bar\gamma}_B})
{\Delta {q}}_C^{n_1}\right]\right),\label{3bfusion}\\
{{{V}}^{n_1n_2n_3}}&=&{g\over8\pi^{3/2}}
\left(\delta_{n_2n_3}\left[(e^{{\bar\beta}_C{\bar\gamma}_C}
-e^{\beta_B\gamma_B})
{\Delta {\hat q}}_A^{n_1}
+e^{\beta_B\gamma_B}{\Delta {q}}_B^{n_1} 
-e^{{\bar\beta}_C{\bar\gamma}_C}{\Delta {q}}_C^{n_1}\right]\right.\nonumber\\
&&\qquad+\delta_{n_1n_3}\left[(e^{\beta_B\gamma_B}
+e^{{\bar\beta}_C{\bar\gamma}_C}){\Delta {q}}_A^{n_2}
-(2e^{{\bar\beta}_C{\bar\gamma}_C}+e^{\beta_B\gamma_B})
{\Delta {\hat q}}_B^{n_2}
+e^{{\bar\beta}_C{\bar\gamma}_C}{\Delta {q}}_C^{n_2}\right]\nonumber\\
&&\qquad\left.+\delta_{n_1n_2}\left[(e^{{\bar\beta}_C{\bar\gamma}_C}+
2e^{\beta_B\gamma_B})
{\Delta {\hat q}}_C^{n_3}-(e^{\beta_B\gamma_B}
+e^{{\bar\beta}_C{\bar\gamma}_C}){\Delta {q}}_A^{n_3}
-e^{\beta_B\gamma_B}{\Delta {q}}_B^{n_3}
\right]\right).
\label{3bfission}
\eea
\subsection{Quartic Vertices}
\label{sec2.3}
We quote here the expressions for the complete quartic vertices,
combining those from ``Coulomb'' exchange with those
from the $\Tr[A_i,A_j]^2$ term in the Lagrangian,
manipulated into a form that suggests a concatenation of
two cubics. In Cartesian basis we can write them in the following way:
\bea
\Gamma^{i_1i_2i_3i_4}&=&{g^2 a\over 32m\pi^3}
\left\{\delta_{i_1i_2}\delta_{i_3i_4}
{(p^+_1-p^+_2)(p^+_4-p^+_3)\over(p^+_1+p^+_2)^2}
+\left(\delta_{i_1i_3}\delta_{i_2i_4}
-\delta_{i_1i_4}\delta_{i_2i_3}\right)\right\}\nonumber\\
&&+{g^2 a\over 32m\pi^3}\left\{\delta_{i_2i_3}\delta_{i_1i_4}
{(p^+_1-p^+_4)(p^+_2-p^+_3)\over(p^+_1+p^+_4)^2}
+\left(\delta_{i_1i_3}\delta_{i_2i_4}
-\delta_{i_1i_2}\delta_{i_3i_4}\right)\right\},
\label{gluon-4-ver}
\eea
where as before all $p^+_k$ are taken to flow {\it into} the vertex,
and the rearrangement factor $a/32m\pi^3$ has also been included.
The first line on the right side looks like the exchange of an
$O(D-2)$ singlet and an $O(D-2)$ antisymmetric tensor in the $s$
channel (12,34) and the second line like the same exchanges in the
$t$ channel (23,41). 
We shall discuss how these quartic vertices as well as quartics
that involve fermion legs are produced from pairs
of cubics in Section 4.
\section{Fermion Fields}
\label{fermions}

In this section, we will extend our discussion to the fermion
case and thus to the supersymmetric theories. In particular, we are
interested in the $\cN = 1, 2$ and $\cN=4$ supersymmetric gauge theories.

We begin with the fermion part of the Lagrangian~(\ref{d-qcd})
\be
{\cal L}_{Dirac} = i\trace\left[\psi^\dagger\G^0\G^\m\,(\pa_\m\psi -
i\,g[A_\m,\psi])\right],
\ee
and the Dirac equation is
\bea
\G^\m\,(\pa_\m\psi - ig\,[A_\m,\psi]) =0\;.
\label{dirac-eq}
\eea
On the lightcone one eliminates half of the components of $\psi$
by writing $\psi=\psi^++\psi^-$, with 
$\psi^\pm\equiv \Gamma^\pm\Gamma^\mp\psi/2$, so $\Gamma^\pm\psi^\pm=0$.
Then the equation for $\psi^+$ doesn't involve time derivatives,
and so these components can be explicitly eliminated:
\bea
\psi^+=-{1\over2\partial_-}\Gamma^+\Gamma^k\left(\partial_k\psi^-
-ig\left[A_k,\psi^-\right]
\right).
\label{psiplus-into-minus}
\eea
In the notation introduced in the appendix, $\psi^+$ consists
of the checked
components $\psi^{\check{a}}$ and $\psi^-$ the unchecked components
$\psi^b$, where the indices $\check{a}, b$ each run over $2^{(D-2)/2}$ values.
Thus we may also express this relation as
\bea
\psi^{\check a} =-{1\over\sqrt{2}\partial_-}
(\G^0\G^k)_{{\check a}b}\left(\partial_k\psi^b
-ig\left[A_k,\psi^b\right]
\right).
\label{psidot-into-nodot}
\eea
In the light-cone gauge, after eliminating $A_+$ and
$\psi^{\check a}$ using Eq.~(\ref{psidot-into-nodot}), 
we obtain the Lagrangian density
\bea
{\cal L} &=&
i\,\trace\left[\psi^{b\,\dagger}\left(\pa_+-\fr{\nabla^2}{2\pa_-}\right)
\psi^b\right]
+\fr{g^2}{2}\,\trace\left(\{\psi^a,\psi^{a\,\dagger}\}
\fr1{\pa_-^2}\{\psi^b,\psi^{b\,\dagger}\}\right)\nn\\
&
&-i\,g^2\,\trace\left(\fr1{\pa_-^2}[\pa_-A_k,A_k]
\{\psi^b,\psi^{b\,\dagger}\}\right)
-\fr{g}{2}\,\trace\left\{\fr{\pa_n}{\pa_-}\psi^{c\,\dagger}(\d^{nk} +
i\,\Sigma^{nk})_{cb}\,[A_k,\psi^b]\right\}\nn\\
& & -\fr{g}{2}\,\trace\left\{[\psi^{c\,\dagger},A_n](\d^{nk}+
i\Sigma^{nk})_{cb}\fr{\pa_k}{\pa_-}\psi^b\right\} +
g\,\trace\left(\psi^{b\,\dagger}
\left[\fr1{\pa_-}\nabla.A,\psi^b\right]\right)\nn\\
& & +\fr{i g^2}{2}\,\trace\left\{[\psi^{c\,\dagger},A_n](\d^{nk}+
\im\Sigma^{nk})_{cb}\fr1{\pa_-}[A_k,\psi^b]\right\},
\label{ferm-enmo-ten-lightcone}
\eea
where we rescaled $\psi\rightarrow 2^{-1/4}\,\psi$ in
Eq.~(\ref{ferm-enmo-ten-lightcone}) and used the identity $\g^n\g^k =
\d^{nk} + i\,\S^{nk}$. 

As before, we expand the free fermion field in momentum modes
\beq
\psi^a(x)=\int {d^{D-1}p\over(2\pi)^{(D-1)/2}}\left(b_a(p)e^{ix\cdot p}
+d^\dagger_a(p)e^{-ix\cdot p}\right) ,
\label{fermion-fourier}
\eeq
where we recall that 
$d^{D-1}p = dp^+d^{D-2}p\theta(p^+)$ and $x\cdot p =
\boldsymbol{x}\cdot\boldsymbol{p} - x^-p^+$.

From Eq.~(\ref{ferm-enmo-ten-lightcone}), one can see that the fermion
free propagator in the mixed $x^+, p^+, \boldsymbol{p}$ representation does
not contribute factors of $1/p^+$ to the vertices, in contrast to the
boson propagators. The lightcone mixed representation propagators
with $p^+>0$ for the particles created by $b^\dagger$ and $d^\dagger$
are
\bea
\theta(x^+-y^+)e^{-i(x^+-y^+)\boldsymbol{p}^2/2p^+},
\qquad -\theta(y^+-x^+)e^{-i(y^+-x^+)\boldsymbol{p}^2/2p^+},
\label{fermion-propagator}
\eea
respectively. To construct a worldsheet system involving
 fermions, the cubic
interactions are again the key ingredients. 
To extract the Feynman rule for a term in the Lagrangian, we first
normal order the term and then assign the operators from
left to right to the legs of the Feynman
diagram going clockwise starting from the top left corner.
This convention of reading off the Feynman rules for generic cubic and
quartic terms in the Lagrangian can be summarized as follows:
\bea
\displaystyle\quad{\trace(A^\dagger B^\dagger C) \to}
{{}\atop\mbox{\psfrag{A}{$A$}\psfrag{B}{$B$}\psfrag{C}{$C$}
\epsfig{file=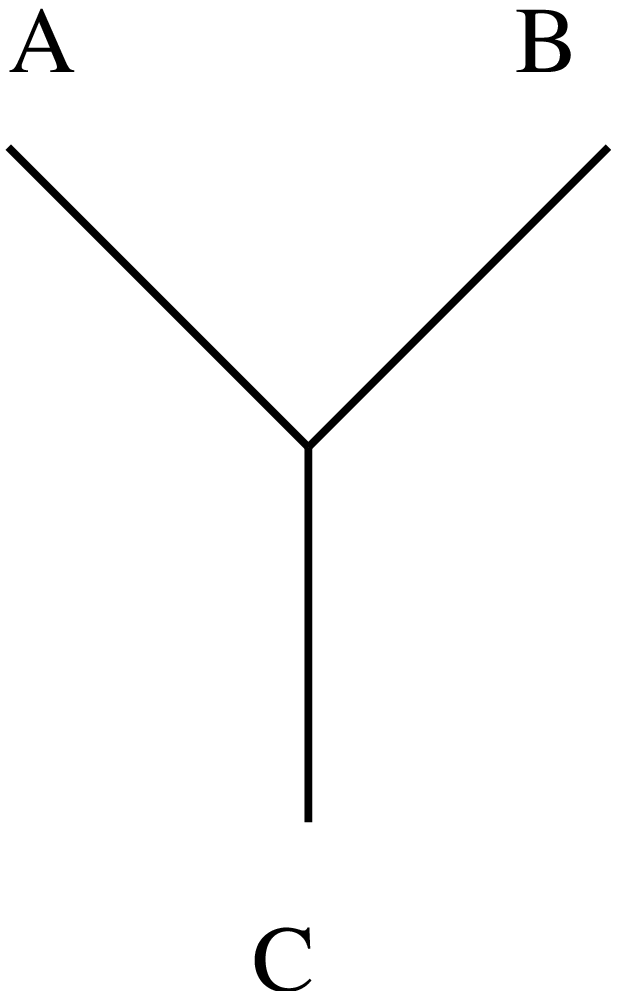,width=1.5cm}}}\;\;\;
\displaystyle\quad{\trace(A^\dagger BC) \to}
{{}\atop\mbox{\psfrag{A}{$A$}\psfrag{B}{$B$}\psfrag{C}{$C$}
\epsfig{file=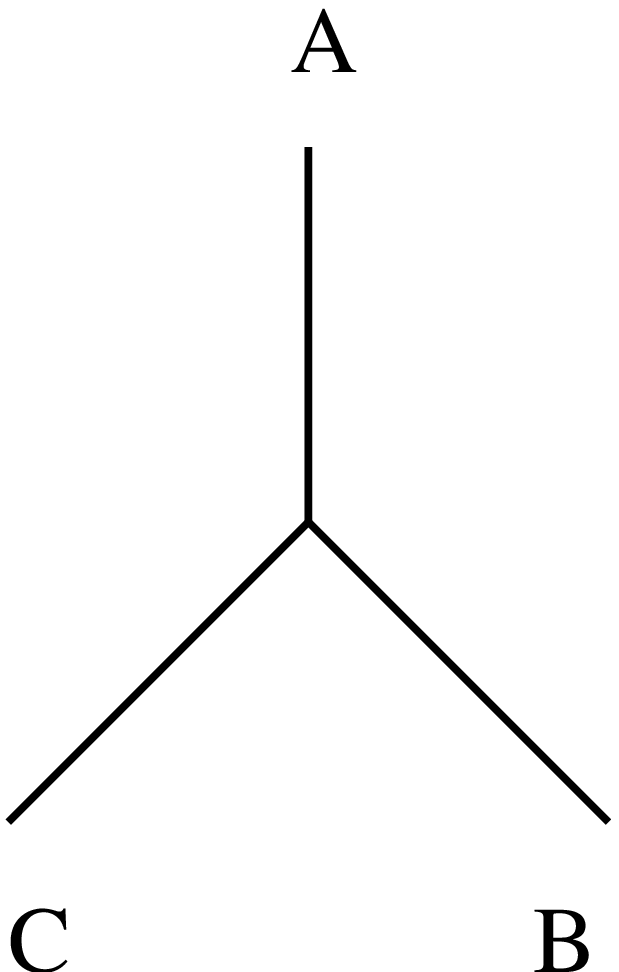,width=1.5cm}}}\;\;\;
\displaystyle\quad{\trace(A^\dagger B^\dagger CD) \to}
{{}\atop\mbox{\psfrag{A}{$A$}\psfrag{B}{$B$}\psfrag{C}{$C$}\psfrag{D}{$D$}
\epsfig{file=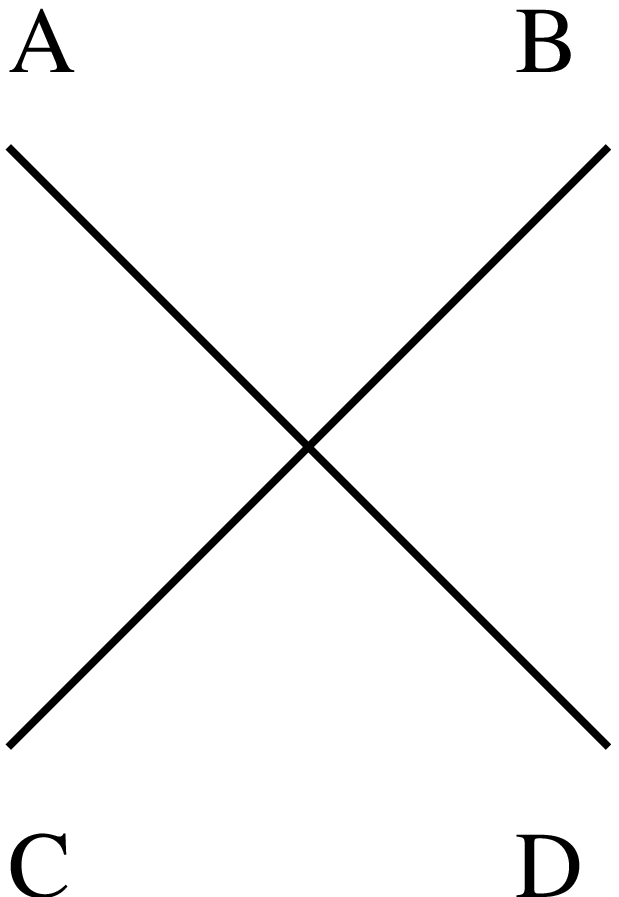,width=1.5cm}}}
\eea

\subsection{Cubic fermion vertices in general dimensions}
\label{cubic-fermions}

\begin{figure}
\begin{center}
\psfrag{'B'}{$b$}
\psfrag{'C'}{$c$}
\psfrag{'n'}{$n$}
\epsfig{file=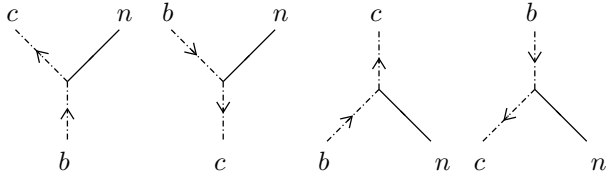,width=8cm}
\caption{Fermion scattering vertices with gluon emitted or
absorbed on the right.} 
\label{fscattr}
\end{center}
\end{figure}

\begin{figure}
\begin{center}
\psfrag{'B'}{$b$}
\psfrag{'C'}{$c$}
\psfrag{'n'}{$n$}
\epsfig{file=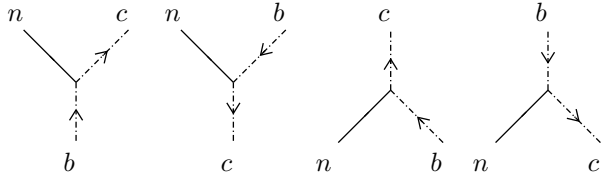,width=8cm}
\caption{Fermion scattering vertices with gluon emitted or
absorbed on the left.} 
\label{fscattl}
\end{center}
\end{figure}

\begin{figure}
\begin{center}
\psfrag{'B'}{$b$}
\psfrag{'C'}{$c$}
\psfrag{'n'}{$n$}
\includegraphics[width=8cm]{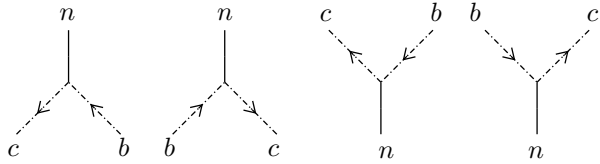}
\caption{Fermion annihilation and creation 
vertices.} 
\label{acvertices}
\end{center}
\end{figure}

Following the analysis in section~\ref{sec2}, we express the cubic term of 
Eq.~(\ref{ferm-enmo-ten-lightcone}) in momentum modes. The general
structure of the cubic term is 
\bea
{\cal L}_1 = \int {dp_bdp_cdp\over\sqrt{p^+}}
\trace\left[(FGF)^k_{cb}+(GFF)^k_{cb}+(FFG)^k_{cb}\right]\,V^k_{cb}\,
\d(p_b+p_c+p),
\label{cubic-hamil}
\eea
where $G$ indicates any of the boson fields, 
$F$ indicates the fermion field and 
\bea
V^{n}_{cb} =
\fr{g}{4\pi^{3/2}}\left[\fr12\,(\g^n\g^k)_{cb}\left(\fr{p_b^k}{p_b^+}-
\fr{p_c^k}{p_c^+}\right)
+ \d_{cb}\left(\fr{p_c^n}{p_c^+}-\fr{p^n}{p^+}\right)\right].
\label{v-def}
\eea
As noted in Section~\ref{sec2}, the coupling constant in
Eq.~(\ref{v-def}) is written in terms of the rescaled coupling
constant appropriate to four-dimensions. Explicit forms of the above mentioned 
structures in Eq.~(\ref{cubic-hamil}) are
\bea
(FFG)^k_{cb} &=&
b_c^\dagger(-p_c)a_k(p)b_b(p_b)+b_c^\dagger(-p_c)a^\dagger_k(-p)b_b(p_b)\nn\\
& & +d_b^\dagger(-p_b)a_k(p)d_c(p_c)+
d_b^\dagger(-p_b)a^\dagger_k(-p)d_c(p_c)\label{group-1}\\ 
(GFF)^k_{cb} &=&
a^\dagger_k(-p)b_c^\dagger(-p_c)b_b(p_b)+a^\dagger_k(-p)
d_b^\dagger(-p_b)d_c(p_c)\nn\\
& &+d^\dagger_b(-p_b)d_c(p_c)a_k(p)+b^\dagger_c(-p_c)b_b(p_b)a_k(p)
\label{group-2} \\
(FGF)^k_{cb} &=& a^\dagger_k(-p)b_b(p_b)d_c(p_c)+
a^\dagger_k(-p)d_c(p_c)b_b(p_b)\nn\\
& & + d_b^\dagger(-p_b)b^\dagger_c(-p_c)a_k(p)+b_c^\dagger(-p_c)d_b^\dagger(-p_b)a_k(p)
\label{group-3}.
\eea
Notice that from
Eq.~(\ref{fermion-propagator}), there is a ($-$) sign in the propagator
for antiparticles, and there is also the overall ($-$) included for
each fermion loop. However in the worldsheet construction we
would like to assign all net relative phases in diagrams to
vertices, to achieve a local description. This can be done
by modifying the above vertex assignments depending on which
fermions are in initial and/or final states. The following
scheme does the job. First, the vertices given above are
taken for the case of particle scattering, when $b$ is in the initial state
and $c$ is in the final state. For antiparticle scattering,
when $c$ is in the initial state and $b$ in the final state,
an extra minus sign is applied. The net effect of these
two modifications is that particles and antiparticles are
seen to couple to gluons in exactly the same way. Namely
the vertices in Fig.~\ref{fscattr}, which correspond to Eq.~(\ref{group-1}),
are assigned the factor $+V^{n}_{cb}$;
and the vertices in Fig.~\ref{fscattl}, corresponding to Eq.~(\ref{group-2}), 
are assigned the factor $-V^{n}_{cb}$. For the annihilation and 
creation vertices, an extra minus
sign is applied to the ``counterclockwise'' circulation of
arrows. This means that all of these vertices, shown in
Fig.~\ref{acvertices} and corresponding to Eq.~(\ref{group-3}) 
are assigned the factor $-V^{n}_{cb}$.
With these modified vertex assignments all positive
$p^+$ propagators are positive and there are no extra $-$
signs for fermion loops. This is the desired local assignment
of phases arising from Fermi statistics.

Now we show how to set up the worldsheet
system corresponding to this theory with interacting fermions and
bosons (including both scalars and gluons). 
Using the notation of subsection~\ref{sec2.1}, 
the cubic vertex factor is
\bea
V^k_{cb} =
\fr{g}{4\p^{3/2}}\,\left[\fr12\,(\g^n\g^k)_{cb}\left(\D\hat{q}_b^k
  -\D\hat{q}_c^k\right) + \d_{cb}\d^{nk}\left(\D\hat{q}_c^k-\D\hat{q}^k
\right)\right].
\label{cubic-ferm-insertion}
\eea
Two points are worth comment. First of all, one
can see from Eq.~(\ref{cubic-ferm-insertion}) that there is no factor of
$M_i$, so one might have thought that one would not need 
to use the $\b, \g$ ghosts. 
However, as we shall
see later in section 5 we do need $\b, \g$ ghosts to compensate
factors which are produced by $b, c$ ghost insertions,
which are applied uniformly to all vertices. Secondly, the cubic
vertex factor in
Eq.~(\ref{cubic-ferm-insertion}) involves only $\D\hat{q}$. As we shall
see later in section~\ref{cubics-to-quartics}, this choice is
necessary to produce the correct quartic vertices.

\subsection{Fermion quartic vertices}
\label{quartic-fermion}

\begin{figure}
\begin{center}
\psfrag{A}{$a$}
\psfrag{B}{$b$}
\psfrag{C}{$c$}
\psfrag{D}{$d$}
\psfrag{I}{${\rm I)}$}
\psfrag{II}{${\rm II)}$}
\psfrag{III}{${\rm III)}$}
\psfrag{IV}{${\rm IV)}$}
\includegraphics[width=10cm]{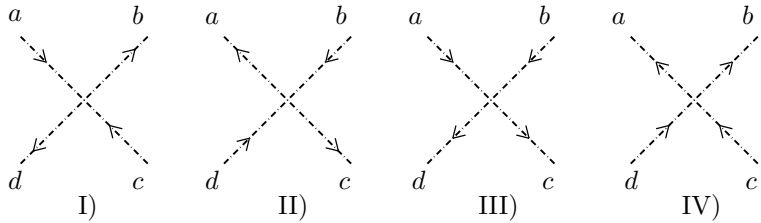}
\caption{Four fermion quartic vertices.} 
\label{4f4l}
\end{center}
\end{figure}

\begin{figure}
\begin{center}
\psfrag{m}{$n_1$}
\psfrag{B}{$b$}
\psfrag{C}{$c$}
\psfrag{n}{$n_2$}
\psfrag{I}{${\rm I)}$}
\psfrag{II}{${\rm II)}$}
\psfrag{III}{${\rm III)}$}
\psfrag{IV}{${\rm IV)}$}
\includegraphics[width=10cm]{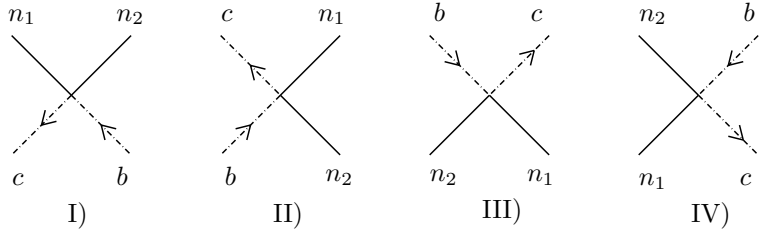}
\caption{Two fermion and two gluon quartic vertices - 1.} 
\label{4lggff1}
\end{center}
\end{figure}

\begin{figure}
\begin{center}
\psfrag{m}{$n_1$}
\psfrag{B}{$b$}
\psfrag{C}{$c$}
\psfrag{n}{$n_2$}
\psfrag{I}{${\rm I)}$}
\psfrag{II}{${\rm II)}$}
\psfrag{III}{${\rm III)}$}
\psfrag{IV}{${\rm IV)}$}
\includegraphics[width=10cm]{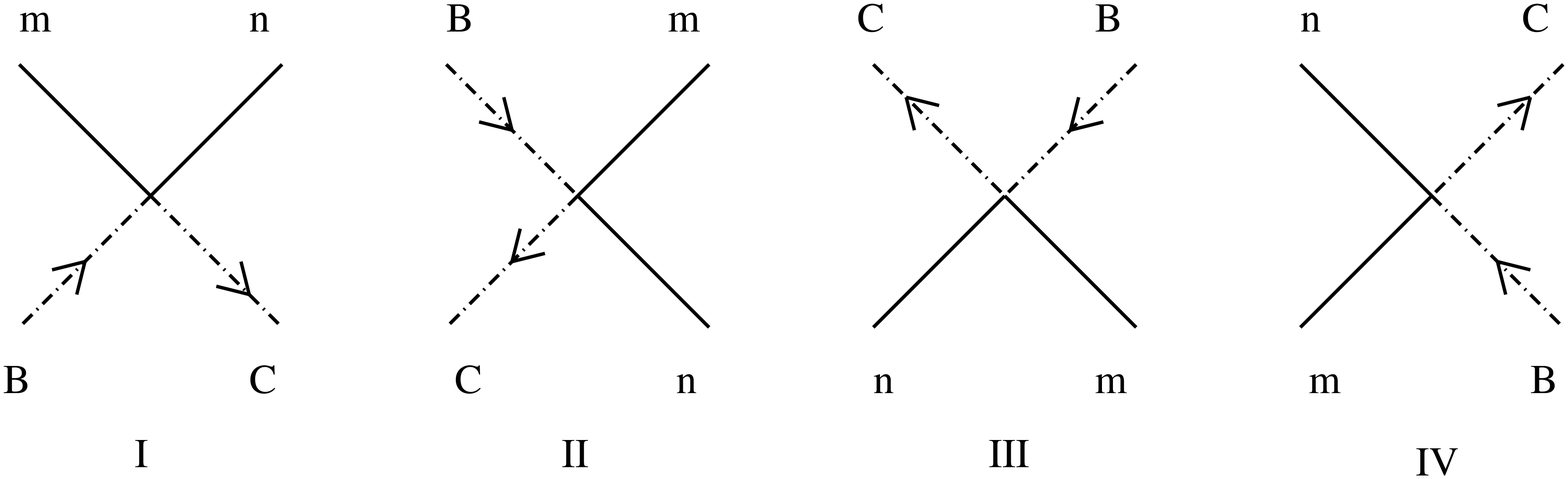}
\caption{Two fermion and two gluon quartic vertices - 2.} 
\label{4lggff2}
\end{center}
\end{figure}

\begin{figure}
\begin{center}
\psfrag{m}{$n_1$}
\psfrag{B}{$b$}
\psfrag{C}{$c$}
\psfrag{n}{$n_2$}
\psfrag{I}{${\rm I)}$}
\psfrag{II}{${\rm II)}$}
\psfrag{III}{${\rm III)}$}
\psfrag{IV}{${\rm IV)}$}
\includegraphics[width=10cm]{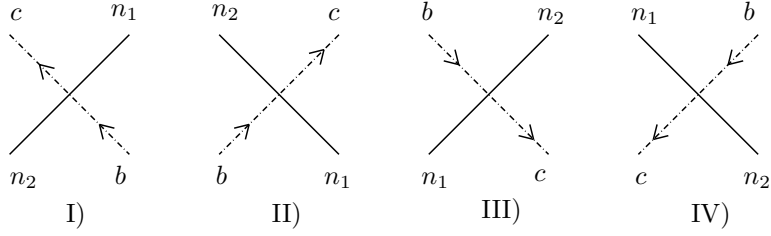}
\caption{Two fermion and two gluon quartic vertices - 3.} 
\label{4lfgfg}
\end{center}
\end{figure}

We present here a complete list of fermion quartic vertices derived
from (\ref{ferm-enmo-ten-lightcone}). 
There are two groups of quartic
vertices. One, which involves four fermions, is shown in
Fig.~\ref{4f4l} and the other, which has two fermion legs and
two gluon legs, is depicted in Fig.~\ref{4lggff1}, Fig.~\ref{4lggff2}
and Fig.~\ref{4lfgfg}. among the
four fermion quartic vertices, there are two types of quartic
couplings: the first group is shown in Fig.~\ref{4f4l}-I)
and~\ref{4f4l}-II), and the other is shown in
Fig.~\ref{4f4l}-III) and~\ref{4f4l}-IV). 
Each of those shown in Fig.~\ref{4f4l}-I) and \ref{4f4l}-II), has the 
coupling
\bea
-\fr{ag^2}{8m\p^3}\,\left[\fr1{(p_a^++p_b^+)^2}\d_{ab}\d_{cd}
-\fr1{(p_a^++p_d^+)^2}\d_{ad}\d_{bc}\right],
\label{coup-4f4l-12}
\eea
whereas each shown in
Fig.~\ref{4f4l}-III) and \ref{4f4l}-IV) has the coupling
\bea
\fr{ag^2}{8m\p^3}\,\fr1{(p_a^++p_d^+)^2}
\d_{ad}\d_{bc}\;.
\label{coup-4f4l-34}
\eea
As one can see from the Lagrangian, there are several types of
two fermion--two gluon quartic couplings. Each of the 
first type, shown in Fig.~\ref{4lggff1}-I), Fig.~\ref{4lggff1}-III) 
and  Fig.~\ref{4lggff1}-IV), has coupling
\bea 
\fr{ag^2}{32m\p^3}\,\left[2\fr{p_1^+-p_2^+}{(p_1^++p_2^+)^2}
\d^{n_1n_2}\d_{cb}+\fr{(\g^{n_1}\g^{n_2})_{cb}}{p_2^++p_b^+}
    \right],
\label{coup-4lggff1-123}
\eea
whereas each shown in Fig.~\ref{4lggff1}-II) have a coupling
\bea 
-\fr{ag^2}{32m\p^3}\,\left[2\fr{p_1^+-p_2^+}{(p_1^++p_2^+)^2}
\d^{n_1n_2}\d_{cb}+\fr{(\g^{n_1}\g^{n_2})_{cb}}{p_2^++p_b^+}
    \right].
\label{coup-4lggff1-4}
\eea
Another type of quartic vertices is shown in Fig.~\ref{4lggff2}. In
this case each of the vertices in 
Fig.~\ref{4lggff2}-I), Fig.~\ref{4lggff2}-III) and
Fig.~\ref{4lggff2}-IV) has the same coupling
\bea
\fr{ag^2}{32m\p^3}\,\left[2\fr{p_1^+-p_2^+}{(p_1^++p_2^+)^2}
\d^{n_2n_1}\d_{cb}-\fr{(\g^{n_2}\g^{n_1})_{cb}}{p_1^++p_b^+}
    \right],
\label{coup-4lggff2-123}
\eea
and Fig.~\ref{4lggff2}-II) has the coupling
\bea
-\fr{ag^2}{32m\p^3}\,\left[2\fr{p_1^+-p_2^+}{(p_1^++p_2^+)^2}
\d^{n_2n_1}\d_{cb}-\fr{(\g^{n_2}\g^{n_1})_{cb}}{p_1^++p_b^+}
    \right].
\label{coup-4lggff2-4}
\eea
The last type of two gluon-two fermion quartic vertices is 
depicted in Fig.~\ref{4lfgfg}. The coupling of Fig.~\ref{4lfgfg}-I) or
Fig.~\ref{4lfgfg}-II) is
\bea
\fr{ag^2}{32m\p^3}\,\left[\fr{(\g^{n_1}\g^{n_2})_{cb}}
{p_2^++p_b^+}+\fr{(\g^{n_2}\g^{n_1})_{cb}}{p_1^++p_b^+}
    \right],
\label{coup-4lfgfg-12}
\eea
while the coupling of Fig.~\ref{4lfgfg}-III) or Fig.~\ref{4lfgfg}-IV)
is just 
\bea
-\fr{ag^2}{32m\p^3}\,\left[\fr{(\g^{n_1}\g^{n_2})_{cb}}
{p_2^++p_b^+}+\fr{(\g^{n_2}\g^{n_1})_{cb}}{p_1^++p_b^+}
    \right].
\label{coup-4lfgfg-34}
\eea
In Section 4, we shall discuss how these quartic vertices are
correctly produced from pairs of boson and fermion cubic vertices.

\section{Quartics from Cubics}
\label{cubics-to-quartics}

In this section we show that the worldsheet mechanism for generating
quartics from pairs of cubics identified in
Ref.~\cite{thornsheet} applies also to the more general theories discussed
here. 

Recall from~\cite{thornsheet} that the scheme
for spreading the propagator among $M$ bits, involved the integral
\begin{eqnarray}
I=\int d^2q_1\cdots d^2q_{M-1}\exp\left\{-{a\over2m}\sum_{i=0}^{M-1}
(\boldsymbol{q}_{i+1}-\boldsymbol{q}_{i})^2\right\}
\equiv\int Dq e^{-S_q}.
\end{eqnarray}
and introduce the shorthand $\langle \cdots\rangle
=\int Dq (\cdots)e^{-S_q}/I$. Then we recall from \cite{thornsheet}
the identities
\begin{eqnarray}
\langle[\boldsymbol{q}_l-\boldsymbol{q}_{l-1}]\rangle 
&=&{\boldsymbol{q}_M-\boldsymbol{q}_0\over M}\\
\left\langle (q^\a_{i+1}-q^\a_i)(q^\b_{j+1}-q^\b_j)\right\rangle
&=&\left({q_M-q_0\over M}\right)^\a\left(\fr{q_M-q_0}{M}\right)^\b
+\fr{m}{a}\d^{\a\b}\d_{ij}-\fr{m}{a}\,\fr{\d^{\a\b}}{M}\; .
\label{master-formula}
\eea
The fluctuation terms of (\ref{master-formula}) 
cause two coincident cubics to behave
as a quartic contact vertex.
Because the
$\Delta\boldsymbol{q}$ insertions on the three different
worldsheet strips joined at a vertex
are applied on two different time slices, we
have the three possible fluctuation 
contributions shown in Fig.~\ref{contact231}.
The a) and b) contributions lead to the
quartic vertices required by the Feynman
rules. However, the c) contribution is 
eliminated by the double ghost insertion
on one of the strips entering the vertex. 
\begin{figure}[ht]
\begin{center}
${\epsfig{file=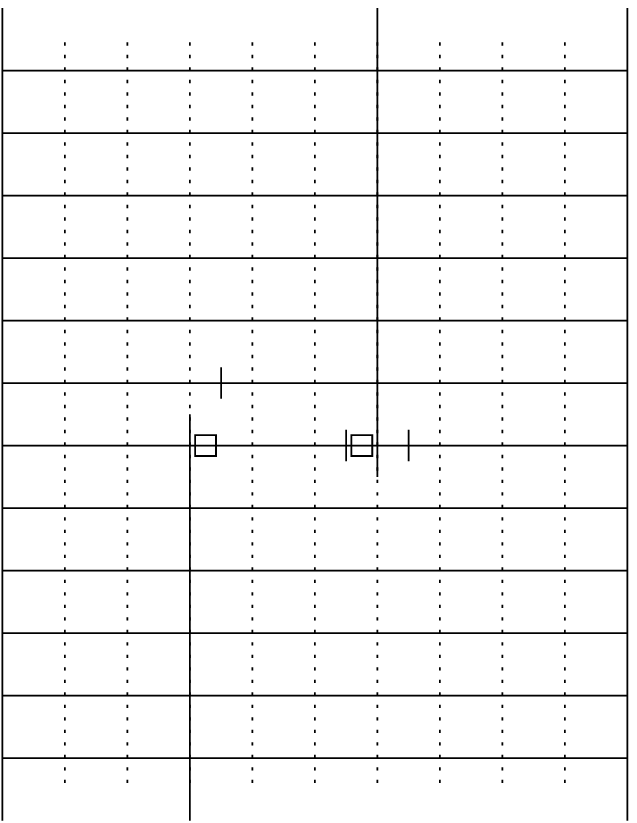,width=4cm}\atop {\displaystyle\rm a)}}
\hskip1cm\quad
{\epsfig{file=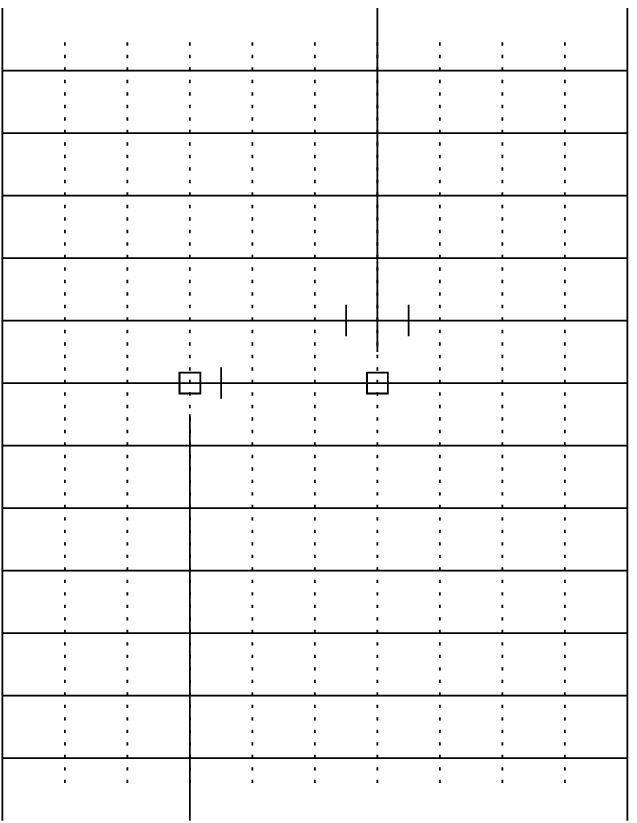,width=4cm}\atop {\displaystyle\rm b)}}
\hskip1cm\quad
{\epsfig{file=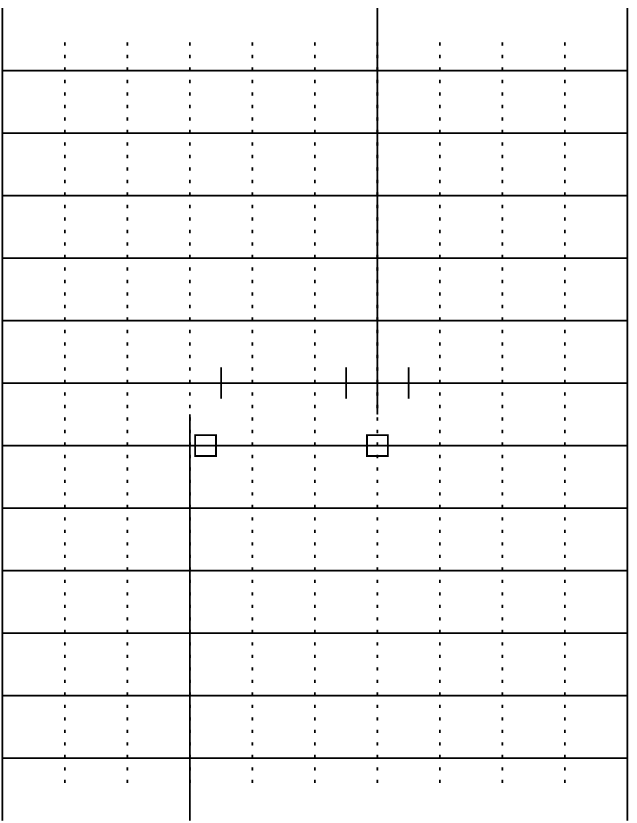,width=4cm}\atop {\displaystyle\rm c)}}
$
\end{center}
\caption{Possible contributions to the quantum term from two
$\Delta\boldsymbol{q}$ insertions placed at the location of the
open squares on the same time slice. Figures a) and b) produce
the desired quartic vertices. 
The tick marks identify the ghost vertex insertions.
The double insertion on strip 4 in figure c) provides a zero that
suppresses this spurious quartic contribution.}
\label{contact231}
\end{figure}
In the next subsections, we shall discuss how to combine two cubic vertices
into a quartic vertex.

\subsection{Boson quartic vertex from two boson cubic vertices}
\label{4-glu-from-2-cu-glu}

In this subsection, we basically repeat arguments in
Ref.~\cite{thornsheet} that a pair of
cubic can combine to correctly produce a quartic vertex.
However in the cases discussed here, 
in addition to gluons there are scalars as a
consequence of dimensional reduction from higher dimensional
theory. Therefore, we have here not only gluon exchange diagrams but also
scalar exchange diagrams. 
 
To see how a pair of two cubic vertices can correctly produce a
quartic vertex, let us consider the four boson diagrams built from two cubics
in Fig.~\ref{fourgluons}. Using Eqs.~(\ref{fissioninsert}) and
(\ref{fusioninsert}), the product of two cubic vertex factors is
\newpage
\bea
&& {g^2a^2\over
  64m^2\pi^3}\left[\d_{n_1n_2}\d_{n_3n_4}(M_1-M_2)(M_3-M_4)
\D\hat{q}^{n_5}\D\hat{q}^{n_5}\right.\nn\\
&&  +
\d_{n_1n_4}(M_1+M_2)(M_3+M_4)\D q^{n_2}\D q^{n_3}
+\d_{n_2n_3}(M_1+M_2)(M_3+M_4)\D q^{n_1}\D q^{n_4}\nn\\
& &\left.-\d_{n_1n_3}(M_1+M_2)(M_3+M_4)\D q^{n_2}\D q^{n_4}
-\d_{n_2n_4}(M_1+M_2)(M_3+M_4)\D q^{n_1}\D q^{n_3}
\right]\nn\\
&& + {g^2a^2\over 64m^2\pi^3}\left[\d_{n_2n_3}\d_{n_1n_4}(M_1-M_4)(M_3-M_2)
\D\hat{q}^{n_5}\D\hat{q}^{n_5}\right.\nn\\
&& +
\d_{n_1n_2}(M_1+M_4)(M_3+M_2)\D q^{n_4}\D q^{n_3}
+\d_{n_4n_3}(M_1+M_4)(M_3+M_2)\D q^{n_1}\D q^{n_2}\nn\\
& &\left.-\d_{n_1n_3}(M_1+M_4)(M_3+M_2)\D q^{n_2}\D q^{n_4}
-\d_{n_2n_4}(M_1+M_4)(M_3+M_2)\D q^{n_1}\D q^{n_3}
\right] .
\label{product-2cubics}
\eea
In this process fluctuation contributions of the
type shown in Fig.~\ref{contact231}a) and Fig.~\ref{contact231}b) come from
a double insertion on the intermediate string
with momentum either $p_1+p_2 = -p_3-p_4$ or $p_1+p_4=-p_2-p_3$.
Remembering the $1/|M_1+M_2|$ (or $1/|M_1+M_4|$) factor from the
intermediate propagator,
the contribution of the quantum term is
\bea
& & \langle\D q^{n_1}\D q^{n_2}\rangle =
-\fr{g^2a}{32m\pi^3}\fr{\d^{n_1n_2}}{|M_1+M_2|^2}\;\rightarrow 
\sum_{n}\langle\D q^{n}\D q^{n}\rangle =
-\fr{g^2a}{32m\pi^3}\fr{(D-2)}{|M_1+M_2|^2},\label{quantumterm-full}\\
& &\langle\D\hat{q}^{n_5}\D\hat{q}^{n_5}\rangle =
-\fr{g^2a}{32m\pi^3}\fr{2}{|M_1+M_2|^2}.
\label{quantumterm-2}
\eea
Note that the factor of 2 in Eq.~(\ref{quantumterm-2}) corresponds to
the two transverse degrees of freedom of gluons (four dimensional
gauge bosons). Eq.~(\ref{quantumterm-full})
yields precisely the contribution of the commutator squared term in the
Yang-Mills Lagrangian (see Eq.~(\ref{d-qcd})), while
Eq.~(\ref{quantumterm-2}) produces the quartic vertex contribution
from the induced instantaneous ``Coulomb'' exchange.
\begin{figure}
\begin{center}
\psfrag{k}{$n_1$}
\psfrag{l}{$n_2$}
\psfrag{m}{$n_3$}
\psfrag{n}{$n_4$}
\psfrag{p}{$n_5$}
\includegraphics[width=7cm]{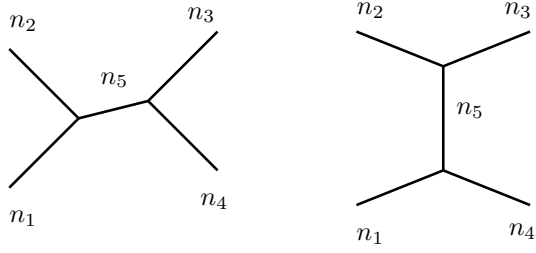}
\caption{Gluon quartic vertex from two cubic vertices.}
\label{fourgluons}
\end{center}
\end{figure}
Substituting Eqs.~(\ref{quantumterm-full}) and~(\ref{quantumterm-2}) 
into Eq.~(\ref{product-2cubics})
and using momentum conservation in $D=4$, it is not hard to show
that the a) and b) type contributions correctly produce the quartic
vertex in Eq.~(\ref{gluon-4-ver}). 

\subsection{Fermion quartic vertices from two cubic vertices}
\label{4-ferm-2-cu-ferm}

Similarly to the previous subsection we shall show how two fermion
cubic vertices also produce the quartic vertices. In principle, two cubic
vertices should exchange either scalars, gluons or fermions to form
quartic vertices. As we already saw in the boson case, to
correctly produce the boson quartic vertex we needed both gluon and scalar
exchanges. In the fermion case, besides gluon exchange we also need 
fermion exchange but not scalar exchange.
As before, the main fluctuation contributions are of the types shown in
Fig.~\ref{contact231}a) and Fig.~\ref{contact231}b) while the unwanted
contribution in Fig.~\ref{contact231}c) due to zero propagation time
for either gluon exchange or fermion exchange between two cubic
vertices is eliminated using the $b, c$ ghost insertion scheme.
There are three distinct cases in which two cubic vertices form a quartic
vertex. The first one, which involves gluon exchange, is shown in
Fig.~\ref{gluon-exchange}. Another process, which involves both gluon and
fermion exchange, is depicted in Fig.~\ref{ferm-glu-exchange}. The
last one shown in Fig.~\ref{fermion-exchange} involves only
fermion exchange.

For the gluon exchange diagrams, there are two different contributions
from $s$- and $t$-channels. The product of two cubic vertices of the
first diagram (\ie $s$-channel) in Fig.~\ref{gluon-exchange} is
\bea
-\fr{a^2g^2}{16m^2\p^3}&\times&
\left[\fr12\,(\g^n\g^k)_{ad}(\D\hat{q}_d^k-\D\hat{q}_a^k)
+ \d_{ad}\d^{nk}(\D\hat{q}^k_a-\D\hat{q}^k)\right]\nn\\
&\times&\left[
\fr12\,(\g^n\g^l)_{cb}(\D\hat{q}_b^l-\D\hat{q}_c^l)
+ \d_{cb}\d^{nl}(\D\hat{q}^l_b-\D\hat{q}^l)\right].
\label{2cubic-s-channel}
\eea
\begin{figure}
\begin{center}
\psfrag{A}{$d$}
\psfrag{B}{$a$}
\psfrag{C}{$b$}
\psfrag{D}{$c$}
\psfrag{n}{$n$}
\includegraphics[width=7cm]{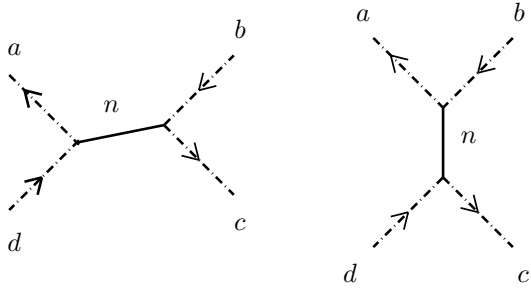}
\caption{Four fermion quartic vertex from two cubic vertices exchange gluons.}
\label{gluon-exchange}
\end{center}
\end{figure}
Notice that the only contribution in Eq.~(\ref{2cubic-s-channel}) comes
from gluons. So one can write Eq.~(\ref{2cubic-s-channel}) as
\bea
-\fr{a^2g^2}{16m^2\p^3}\left[ \d_{ad}\d_{bc}\D\hat{q}^n\D\hat{q}^n\right],
\eea
and the quantum fluctuation term gives
\bea
\langle\D\hat{q}^n\D\hat{q}^n\rangle =
-\fr{m}{a}\fr{2}{(p_a^++p_d^+)^2}.
\eea
So the contribution from $s$-channel gives us
\bea
\fr{ag^2}{8m\pi^3}\,\fr{\d_{ad}d_{bc}}{(p_a^++p_d^+)^2},
\label{t-channel-res}
\eea
where we already included the factor $1/|M_1+M_4|$ from the gluon
intermediate propagator. Similarly, one obtains for the
$t$-channel
\bea
-\fr{ag^2}{8m\pi^3}\,\fr{\d_{ab}d_{cd}}{(p_a^++p_b^+)^2}.
\label{s-channel-res}
\eea
Combining Eqs.~(\ref{t-channel-res}) and (\ref{s-channel-res}), it is
easy to confirm that two cubic vertices indeed produce the four fermion
quartic vertices shown in Fig.~\ref{4f4l}-II) with
coupling~(\ref{coup-4f4l-12}). Notice that in this case only 
the instantaneous
``Coulomb'' exchanges contribute. Other diagrams of this type can be
analyzed in the same way. Note that to produce diagrams
Fig.~\ref{4f4l}-III) and~\ref{4f4l}-IV) one needs only the $s$-channel
contribution. 

For quartic vertices involving two fermions and two gluons, there are
three different kinds of diagrams as shown in Figs.~\ref{4lggff1},
\ref{4lggff2} and \ref{4lfgfg}. The first set of quartic vertices in
Fig.~\ref{4lggff1} can be constructed from two cubic vertices which
exchange both gluon and fermion as shown in Fig.~\ref{ferm-glu-exchange}.
\begin{figure}
\begin{center}
\psfrag{m}{$n_2$}
\psfrag{n}{$n_1$}
\psfrag{p}{$n_3$}
\psfrag{B}{$b$}
\psfrag{C}{$c$}
\psfrag{D}{$d$}
\includegraphics[width=7cm]{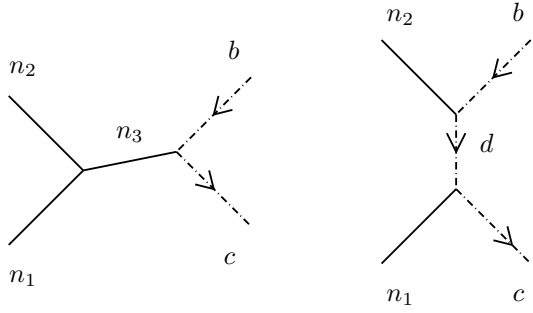}
\caption{Quartic vertices involving fermions and gluons exchange both
  gluons and fermions.}
\label{ferm-glu-exchange}
\end{center}
\end{figure}
In this
case the product of two cubic vertices from gluon exchange is
\bea
&&-\fr{a^2g^2}{32m^2\p^3}\left(\d^{n_1n_2}\left[-(M_3+M_2)\D
  q_A^{n_3}+(2M_2+M_3)\D q_B^{n_3} - M_2\D\hat{q}_C^{n_3}\right]\right.
\nn\\
&&+\d^{n_1n_3}\left[(M_3+M_2)\D q_A^{n_2} +
M_3\D\hat{q}_B^{n_2} - (M_2+2M_3)\D q_C^{n_2}\right]\nn\\
&&+\left.\d^{n_2n_3}\left[(M_3-M_2)\D\hat{q}_A^{n_1}-M_3\D q_B^{n_1}
 +M_2\D q_C^{n_1}\right]\right)\nn\\
&&\times
\left[\fr12(\g^{n_3}\g^k)_{cb}(\D\hat{q}_b^k-\D\hat{q}_c^k) + \d_{cb}\d^{n_3k}(\D\hat{q}_c^k-\D\hat{q}^k)\right].
\label{ccc}
\eea
Because this is a gluon exchange diagram, the important terms 
are those hat momenta with upper index $n_3$, \ie Eq.~(\ref{ccc}) can
be written as
\bea
+\fr{a^2g^2}{32m^2\p^3}\d^{n_1n_2}\d_{cb}(M_3+2M_1)
\D\hat{q}^{n_3}\D\hat{q}^{n_3} + \cdots\; .
\eea
Using our master 
formula~(\ref{master-formula}) and $M_2=-M_1-M_3$, the above equation gives
\bea
\fr{ag^2}{16m\p^2}\,\fr{p_1^+-p_2^+}{(p_1^++p_2^+)^2}\,\d^{n_1n_2}\d_{cb} .
\label{fgx-gexchange}
\eea
On the other hand, the product of two cubic vertices with fermion
exchange shown in Fig.~\ref{ferm-glu-exchange} is
\bea
&&\fr{a^2g^2}{16m^2\p^2}\left[\fr12(\g^{n_1}\g^k)_{cd}(\D\hat{q}_d^k-\D\hat{q}_c^k)
+ \d_{cd}(\D\hat{q}_c^{n_1}-\D\hat{q}^{n_1})\right]\nn\\
&&\times
\left[\fr12(\g^l\g^{n_2})_{db}(\D\hat{q}_d^l-\D\hat{q}_b^l) +
\d_{db}(\D\hat{q}_b^{n_2}-\D\hat{q}^{n_2})\right].
\eea
In this case the contribution comes from the hat momenta carrying
spinor index $d$. Using our master formula~(\ref{master-formula}) and
momentum conservation $M_d = p_1^++p_c^+ = -(p_2^++p_b^+)$ (notice
that $M_d>0$) we obtain
\bea
\fr{ag^2}{32m\p^3}\,\fr{(\g^{n_1}\g^{n_2})_{cb}}{p_2^++p_b^+}.
\label{fgx-fexchange}
\eea
Combining Eqs.~(\ref{fgx-gexchange}) and (\ref{fgx-fexchange}) one
gets
\bea
\fr{ag^2}{32m\p^2}\left[2\fr{p_1^+-p_2^+}{(p_1^++p_2^+)^2}\,\d^{n_1n_2}\d_{cb}
+\fr{(\g^{n_1}\g^{n_2})_{cb}}{p_2^++p_b^+}\right].
\eea
The above equation is precisely a coupling of 
quartic vertex shown in Fig.~\ref{coup-4lggff1-4}-IV).

Similarly, we can show that the other set of quartic vertices
involving two fermions and two gluons as depicted in Fig.~\ref{4lfgfg} can
be correctly reproduced by combining two cubic vertices. In this
case there are two contributions coming from $s$- and $t$-channels
of fermion exchange diagrams as shown Fig.~\ref{fermion-exchange}.
\begin{figure}
\begin{center}
\psfrag{m}{$n_1$}
\psfrag{n}{$n_2$}
\psfrag{C}{$c$}
\psfrag{D}{$d$}
\psfrag{B}{$b$}
\includegraphics[width=7cm]{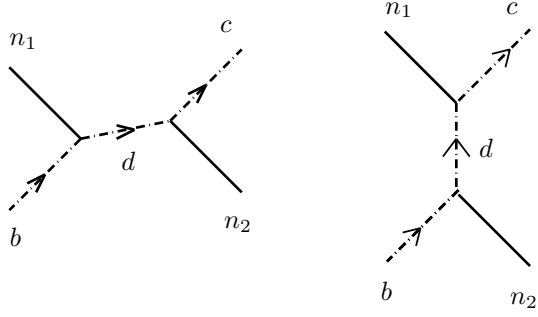}
\caption{Quartic vertices involving fermions and gluons exchange fermions.}
\label{fermion-exchange}
\end{center}
\end{figure}
The product of two cubic vertices of this kind is
\bea
&&-\fr{a^2g^2}{16m^2\p^2}\left\{\left[\fr12(\g^{n_2}\g^k)_{db}(\D\hat{q}_b^k-\D\hat{q}_d^k)
+ \d_{db}(\D\hat{q}_d^{n_2}-\D\hat{q}^{n_2})\right]\times \right.\nn\\
&&\times
\left[\fr12(\g^{n_1}\g^l)_{cd}(\D\hat{q}_d^l-\D\hat{q}_c^l) +
\d_{cd}(\D\hat{q}_c^{n_1}-\D\hat{q}^{n_1})\right]\nn\\
&&+\left[\fr12(\g^{n_1}\g^k)_{db}(\D\hat{q}_b^k-\D\hat{q}_d^k) +
\d_{db}(\D\hat{q}_d^{n_1}-\D\hat{q}^{n_1})\right]\times \nn\\
&&\left.\times \left[\fr12(\g^{n_2}\g^l)_{cd}(\D\hat{q}_d^l-\D\hat{q}_c^l)
+ \d_{cd}(\D\hat{q}_c^{n_2}-\D\hat{q}^{n_2})\right]\right\},
\eea
where the first two lines are contributions from the $t$-channel while the
last two are from the $s$-channel. Again using our master 
formula~(\ref{master-formula}) we get
\bea
\fr{ag^2}{32m\p^3}\left[\fr{(\g^{n_1}\g^{n_2})_{cb}}{p_2^++p_b^+}
+\fr{(\g^{n_2}\g^{n_1})_{cb}}{p_1^++p_b^+} \right],
\eea
and the above equation is precisely the
coupling which is shown in Fig.~\ref{4lfgfg}-II).
\section{Grassmann Variables}
\label{sec5}
In the case of pure Yang-Mills theory, the flow of vector polarization
through a large planar diagram was described by Grassmann odd
worldsheet spin variables 
$\boldsymbol{ S}_k$, which carry transverse
vector indices \cite{thornsheet}. But since we must now deal with 
fermions as well, it is reasonable to instead attach spinor indices to the
Grassmann variables. It will turn out that vector valued spin
variables will not then be needed.

We first assume we have a complete Dirac multiplet of
fermions, \ie\ $2^{(D-2)/2}$ particles and the same number of
anti-particles. In terms of the $O(D-2)$ Clifford algebra,
the rotation generators for particles are $\Sigma^{kl}/2$
whereas those of the anti-particles are $-\Sigma^{kl*}/2$.
(For a Majorana representation the two representations
are identical.)
Thus we introduce Grassmann variables
${S}_k^a$ to describe the
particle spin states and ${\bar S}_k^a$ to describe the
anti-particle spin states. Here $a$ is an $O(D-2)$ spinor index and
$k$ labels the location of the
variable on the worldsheet, which will be specified in more detail later.
Let us first consider
a single chain 
$(S^a_1,{\bar S}^a_1),({S}^a_2,{\bar S}^a_2),
\ldots, ({S}^a_{2K},{\bar S}^a_{2K})$,
with an  {\it even} number of spin variables, each carrying a 
spinor index and use the action
\bea
{\cal A}=\sum_{i=1}^{2K-1}{\bar S}^a_i{S}^a_{i+1}
+\sum_{i=1}^{2K-1}{S}^a_i{\bar S}^a_{i+1}.
\eea
We define the measure for integration of the
Grassmann variables by
\bea
{\cal D}S\equiv\prod_{a}\left[dS_{2K}^adS_{2K-1}^a\cdots dS_1^a\right]
\left[d{\bar S}_{2K}^a d{\bar S}_{2K-1}^a\cdots d{\bar S}_1^a\right].
\eea
Then it is easy to check that
\bea
\langle e^{{\bar\eta}^a_1{S}^a_1+{\eta}^a_1{\bar S}^a_1
+{\eta}^a_{2K}{\bar S}^a_{2K}+{\bar\eta}^a_{2K}{S}^a_{2K}}\rangle
&=&
e^{{\bar\eta}^a_{2K}{\eta}^a_1+{\eta}^a_{2K}{\bar\eta}^a_1},
\eea
where $\langle\cdots\rangle\equiv\int{\cal D}S e^{\cal A}(\cdots)$.
In particular, the last equation implies
\bea
\langle S_1^a\ {\bar S}_{2K}^b\rangle&=&\langle {\bar S}_1^a\ {S}_{2K}^b\rangle
=\delta_{ab}\\
\langle {\bar S}_1^a\ {\bar S}_{2K}^b\rangle&=&
\langle {S}_1^a\ {S}_{2K}^b\rangle
=0.
\label{kdeltao}
\eea
Note that these formulas {\it require} an even number of spins.
In order to guarantee the consistent application of the
even spin formal in the presence of interactions, the
number of spins assigned to each bit must be a multiple of four.
Eq.~\ref{kdeltao} will be used to supply the spinor Kronecker
delta's for fermion propagators. To supply the Kronecker
delta's for the vector and scalar particles, we could introduce
vector valued spinors as in \cite{thornsheet}. However, it
will be more convenient to instead employ the
 bilinear 
$J^j\equiv 2^{-(D-2)/4}S_k^{a}\gamma^j_{ab}{\bar S}_k^b$. For this
scheme to work, it is important that it produces the relations
\bea
\langle S_1^a\ J_{2K}^j\rangle&=&
\langle {\bar S}_1^a\ J_{2K}^j\rangle=
\langle J_1^j\ S_{2K}^a\rangle=
\langle J_1^j\ {\bar S}_{2K}^a\rangle=0\\
\langle J_1^i\ J_{2K}^j\rangle&=&\delta_{ij} .
\label{kdelta1}
\eea
The first line is trivially true. The left side of the 
second equation is just $2^{-(D-2)/2}\gamma^i_{ab}\gamma^j_{ba}$,
with repeated indices summed which is just
$2^{-(D-2)/2}\Tr\gamma^i\gamma^j=\delta_{ij}$, as desired. 

The purpose of the Grassmann variables is to give a worldsheet local
formalism to transport the spin and polarization information
from the external lines to the point within the worldsheet
where the interaction occurs. For the propagator we arrange the
spin chain to visit every site on the lattice worldsheet, by
snaking it through as shown in Fig.~\ref{gluonprop}. 
\begin{figure}[ht]
\begin{center}
{\includegraphics[width=4cm]{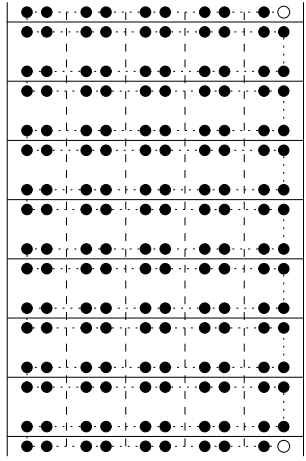}}
\caption{Assignment of Grassmann spins for propagator. Each dot is assigned
Grassmann spins ${S}_k^a,{\bar S}_k^a$,  
and the bond pattern for the spin chain is indicated by dotted lines.
External state information is specified by inserting
${S}_k^a,{\bar S}_k^a$ or the bilinear
$J_k^j$ at the open circles.}
\label{gluonprop}
\end{center}
\end{figure}

\begin{figure}[ht]
\psfrag{'k'}{$k$}
\psfrag{'l'}{$l$}
\centerline{\epsfig{file=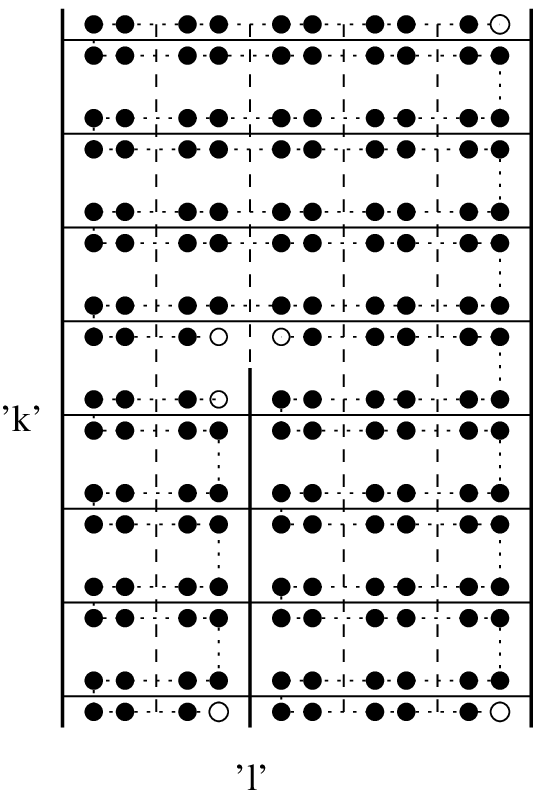,width=5cm}
\hskip1in\epsfig{file=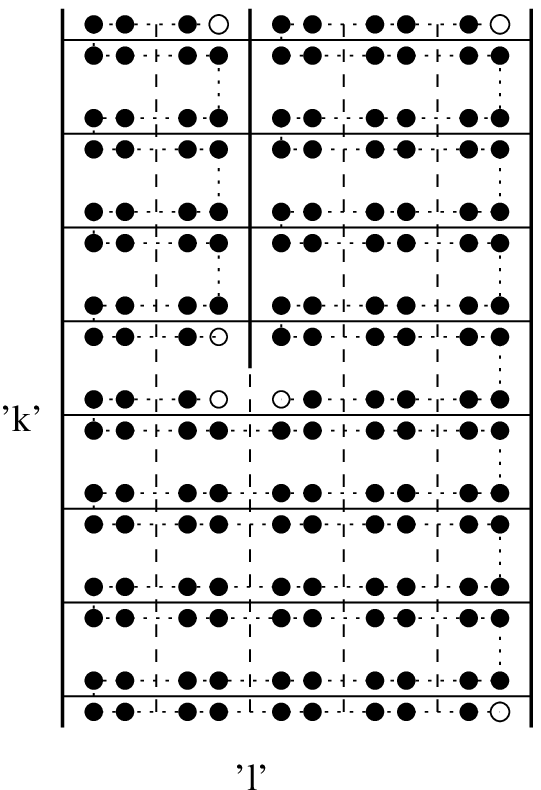,width=5cm}
}
\caption{Cubic fusion and fission vertices. The open circles
surrounding each interaction point are labeled by
the same letters we have used to label the bonds
they occur on in Fig.~\ref{pinserts}.}
\label{cubicvert}
\end{figure} 

Next we turn to interactions. 
We draw the worldsheets for the fusion and fission 
cubic vertices in Fig.~\ref{cubicvert}.
The three open circles, on the bonds we have marked
$A$, $C$, and $D$ in Fig.~\ref{pinserts}, indicate the spins,
which we label by the same letters, 
that participate in the
vertex insertion that is designed to yield the correct cubic vertex. 
Let $k$ label the time slice just before the solid
line ends or begins and let $l$ label the spatial location of the
solid line. In constructing the vertex we need
to refer
to variables on the four spatial links immediately
surrounding the vertex. For this purpose we use
the $A,B,C,D$ labeling scheme indicated in
Fig.~\ref{pinserts}. We must also remember the ghost insertions 
that produce the factors of
$1/|M_i|$ reassigned from the field theoretic bosonic
propagator to the {\it earlier} vertex attached to it. Thus we
insert $e^{-a\Delta b_C\Delta c_C/m}$ in the fusion vertex and
$e^{-a(\Delta b_B\Delta c_B+\Delta b_C\Delta c_C)/m}$ in the fission
vertex. We shall apply these same insertions to vertices involving
fermion lines, even though the fermion propagator had no such
factors of $1/|M_i|$ to begin with. This means that the
vertices involving fermions will include additional factors
of $|M_i|$ in the numerator to cancel the effects of the
ghost insertions. The contribution of the 3 boson vertices
to the insertion is simply a product of three bilinears
$J^k$ times one of the expressions (\ref{3bfusion})
or (\ref{3bfission}). The contribution of the
fermions requires a more elaborate notation because there
are six distinct vertex types. We therefore define the
following three bilocal bilinears:
\bea
{\bar{\cal J}}_{DA}^{j}&=&{g\over8\pi^{3/2}}\big\{S_D^c{\bar S}_A^b
[(\gamma^j\gamma^k)_{cb}(\Delta {\hat q}^k_A-\Delta {\hat q}^k_C)
+2\delta_{cb}(\Delta {\hat q}^k_C-\Delta {\hat q}^k_B)]\nonumber\\
&&+{\bar S}_D^b{S}_A^c[(\gamma^j\gamma^k)_{cb}(\Delta {\hat q}^k_C-\Delta {\hat q}^k_A)
+2\delta_{cb}(\Delta {\hat q}^k_A-\Delta {\hat q}^k_B)]\big\}
(e^{\beta_A\gamma_A}+e^{{\bar\beta}_B{\bar\gamma}_B})\\
{\bar{\cal J}}_{CD}^{j}&=&{g\over8\pi^{3/2}}\big\{S_C^c{\bar S}_D^b
[(\gamma^j\gamma^k)_{cb}(\Delta {\hat q}^k_B-\Delta {\hat q}^k_C)
+2\delta_{cb}(\Delta {\hat q}^k_C-\Delta {\hat q}^k_A)]\nonumber\\
&&+{\bar S}_C^b{S}_D^c[(\gamma^j\gamma^k)_{cb}(\Delta {\hat q}^k_C-\Delta {\hat q}^k_B)
+2\delta_{cb}(\Delta {\hat q}^k_B-\Delta {\hat q}^k_A)]\big\}
(e^{\beta_A\gamma_A}+e^{{\bar\beta}_B{\bar\gamma}_B})\\
{\bar{\cal J}}_{AC}^{j}&=&{g\over8\pi^{3/2}}\big\{S_A^c{\bar S}_C^b
[(\gamma^j\gamma^k)_{cb}(\Delta {\hat q}^k_B-\Delta {\hat q}^k_A)
+2\delta_{cb}(\Delta {\hat q}^k_A-\Delta {\hat q}^k_C)]\nonumber\\
&&+{\bar S}_A^b{S}_C^c[(\gamma^j\gamma^k)_{cb}(\Delta {\hat q}^k_A-\Delta {\hat q}^k_B)
+2\delta_{cb}(\Delta {\hat q}^k_B-\Delta {\hat q}^k_C)]\big\},
\eea
for the fusion case, and a similar set for the fission case:
\bea
{{\cal J}}_{DA}^{j}&=&{g\over8\pi^{3/2}}\big\{S_D^c{\bar S}_A^b
[(\gamma^j\gamma^k)_{cb}(\Delta {\hat q}^k_A-\Delta {\hat q}^k_B)
+2\delta_{cb}(\Delta {\hat q}^k_B-\Delta {\hat q}^k_C)]\nonumber\\
&&+{\bar S}_D^b{S}_A^c[(\gamma^j\gamma^k)_{cb}(\Delta {\hat q}^k_B-\Delta {\hat q}^k_A)
+2\delta_{cb}(\Delta {\hat q}^k_A-\Delta {\hat q}^k_C)]\big\}e^{{\bar\beta}_B{\bar\gamma}_B}
\\
{{\cal J}}_{CD}^{j}&=&{g\over8\pi^{3/2}}\big\{S_C^c{\bar S}_D^b
[(\gamma^j\gamma^k)_{cb}(\Delta {\hat q}^k_A-\Delta {\hat q}^k_C)
+2\delta_{cb}(\Delta {\hat q}^k_C-\Delta {\hat q}^k_B)]\nonumber\\
&&+{\bar S}_C^b{S}_D^c[(\gamma^j\gamma^k)_{cb}(\Delta {\hat q}^k_C-\Delta {\hat q}^k_A)
+2\delta_{cb}(\Delta {\hat q}^k_A-\Delta {\hat q}^k_B)]\big\}e^{\beta_C\gamma_C}\\
{{\cal J}}_{AC}^{j}&=&{g\over8\pi^{3/2}}\big\{S_A^c{\bar S}_C^b
[(\gamma^j\gamma^k)_{cb}(\Delta {\hat q}^k_C-\Delta {\hat q}^k_B)
+2\delta_{cb}(\Delta {\hat q}^k_B-\Delta {\hat q}^k_A)]\nonumber\\
&&+{\bar S}_A^b{S}_C^c[(\gamma^j\gamma^k)_{cb}(\Delta {\hat q}^k_B-\Delta {\hat q}^k_C)
+2\delta_{cb}(\Delta {\hat q}^k_C-\Delta {\hat q}^k_A)]\big\}e^{\beta_C\gamma_C
+{\bar\beta}_B{\bar\gamma}_B}.
\eea
Then the fusion and fission vertex insertions on the worldsheet are
\bea
{{\bar{\cal V}}_l^k\over2\pi}&=&{a\over m}\{{\bar{\cal J}}_{DA}^{j}J_C^j
+{\bar{\cal J}}_{CD}^{j}J_A^j+{\bar{\cal J}}_{AC}^{j}J_D^j
+J_{A}^{n_1}J_{C}^{n_2}J_{D}^{n_3}
{{\bar{V}}^{n_1n_2n_3}}\}e^{-a\Delta b_C\Delta c_C/m}
\label{3barv}\\
{{\cal V}_l^k\over2\pi}&=&{a\over m}\{{{\cal J}}_{DA}^{j}J_C^j
+{{\cal J}}_{CD}^{j}J_A^j+{{\cal J}}_{AC}^{j}J_D^j
+J_{A}^{n_1}J_{C}^{n_2}J_{D}^{n_3}
{{V}}^{n_1n_2n_3}\}e^{-a(\Delta b_B\Delta c_B+\Delta b_C\Delta c_C)/m}.
\label{3v}
\eea
We refer the reader to \cite{thornsheet} for the details of how
these vertex insertions enter the worldsheet path integral
(see especially Eqs.~(24) and (27) of that work).

We now consider briefly how the preceding discussion must be
modified to realize various super symmetries. The number of
fermionic states must be reduced from the complete Dirac multiplet
so far assumed. For ${\cal N}=1,2$ we must reduce the number
by a factor of 2 and for ${\cal N}=4$ by a factor of 4. A factor
of 2 is easily achieved by making a Weyl restriction on the
fermion field. In the basis we have described in the appendix,
this is achieved by restricting the $O(D-2)$ spinor indices on the
insertions $S^a,{\bar S}^a$ to the first (or last) $2^{(D-4)/2}$
components. To distinguish these two sets of
indices, we shall denote the first $2^{(D-4)/2}$ components
by an undotted index $a$ and put a dot over the index, ${\dot a}$
if it labels the second $2^{(D-4)/2}$ components. Thus we start
out with four distinct spinor types 
$S^a,S^{\dot a},{\bar S}^a,{\bar S}^{\dot a}$, and the Weyl
restriction means that we only insert $S^a,{\bar S}^a$ on 
fermion lines. But the boson bilinears require the dotted indices
as well $J^k=2^{-(D-2)/2}(S^{\dot a}\gamma^k_{\dot a b}{\bar S}^b
+S^{a}\gamma^k_{a\dot b}{\bar S}^{\dot b})$, so the 
full complement of spinor indices is retained in the worldsheet
path integral. This procedure takes care of the cases ${\cal N}=1,2$.

The ${\cal N}=4$ case requires a further reduction of fermionic
components by a factor of two. This is allowed because for
$O(8)$ spinors can be made simultaneously Majorana and
Weyl. In the Majorana representation, where $\gamma^k$ are
real (and symmetric) it is consistent to
identify $S$ and ${\bar S}$, so there is no distinction
between particle and anti-particle. But then the bilinear 
$S^a\gamma^k_{a{\dot b}}S^{\dot b}+S^{\dot a}\gamma^k_{\dot a b}S^{b}$
is identically zero. However, precisely in this case
we can redefine the bilinears as 
$J^k\equiv i2^{-(D-4)/2}S^a\gamma^k_{a\dot b}S^{\dot b}$. This
modified definition works because we have
\bea
\langle J_1^k J_{2K}^l\rangle
=-i^22^{-(D-4)/2}\gamma^k_{a\dot b}\gamma^l_{a\dot b}
=2^{-(D-4)/2}\Tr\ \gamma^k\gamma^l{1+\gamma_9\over2}
=\delta_{kl}.
\eea

Finally, we remark that the spinor-valued Grassmann odd variables
we have introduced here can also be employed in the worldsheet
for pure Yang-Mills theory instead of the vector-valued ones
used in \cite{thornsheet}. One simply restricts the insertions
to {\it only} the bilinears $J^k$. We might then dub this
case ${\cal N}=0$ supersymmetry, and the formalism developed
here then covers in a unified manner
all the interesting 4 dimensional quantum
field theories with ${\cal N}=0,1,2,4$ supersymmetry.

\section{One Loop Renormalization}
\label{sec6}

When using the discretized worldsheet to calculate processes to a
given order in perturbation theory, we recall that 
the insertions have been designed to 
exactly reproduce the cubic vertices of the light-cone
Feynman rules in the continuum limit. The precise meaning of this
limit is that every solid line in the diagram is many lattice
steps long and also is many lattice steps away from every other
solid line. Clearly a diagram in which one of these criteria is
not met is sensitive to the details of our discretization choice.
In tree diagrams one can always avoid these dangerous situations
by restricting the external legs so that they carry $p^+_i$
so that all differences $|p^+_i-p^+_j|\gg m$, and so that
the time of evolution between initial and final states $\tau\gg a$.
However, a diagram containing one or more loops will involve
sums over intermediate states that violate these inequalities,
and because of field theoretic divergences the dangerous regions
of these sums can produce significant effects in the
continuum limit. In particular we should expect these effects
to include a violation of Lorentz invariance, in addition to
the usual harmless effects that are absorbed into renormalized
couplings. Indeed, when a solid line is of order a few lattice
steps in length, it produces a gap in the gluon energy spectrum
that is forbidden by Lorentz invariance. This effect can
be canceled by a counter-term that represents a local
modification of the worldsheet action.
We conjecture that all counter-terms needed for
a consistent renormalization program
can be implemented by local modifications of the worldsheet dynamics.
In this section we confirm this
conjecture to one loop order in perturbation theory.

\subsection{Gluon Self Energy}
\label{sec6.1}
The gluon self energy to one loop can be extracted
from the lowest order correction to a gluon propagator
represented by a single
solid line segment on a worldsheet strip as in
Fig.~\ref{oneloopse}. 
\begin{figure}[!ht]
\begin{center}
\psfrag{'l'}{$l$}
\psfrag{'M'}{$M$}
\psfrag{'k0'}{$k_1$}
\psfrag{'k1'}{$k_2$}
\includegraphics[width=6cm]{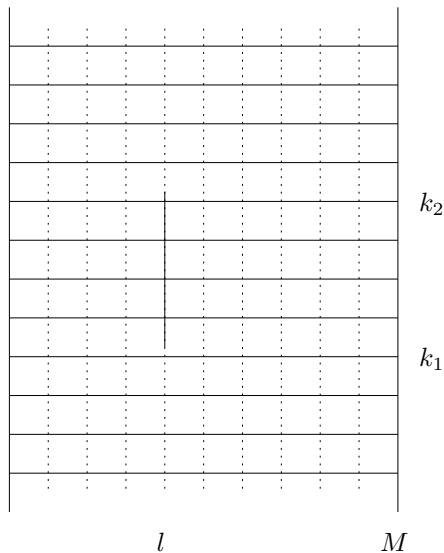}
\caption{One loop self energy calculation. Because of time translation
invariance only the difference $k=k_2-k_1$ is important.}
\label{oneloopse}
\end{center}
\end{figure}
For the theories considered in this article we must add the
contribution of the fermions and scalars to the
calculation in the pure Yang-Mills theory given 
in \cite{beringrt}. With our conventions the result
analogous to Eqs.~(52) and (53) of that article, for fixed $k_1,k_2,l$
with $k=k_2-k_1>1$ and $0<l<M$, reads:
\bea
\Pi_{l,k} =\frac{g^2}{8\pi^2} \frac{u^k}{k^2}
\left\{{2\over l}
+{2\over M-l}-{1\over M}\left[4-{N_f\over2}-(2+N_s-N_f){l\over M}
\left(1-{l\over M}\right)\right]
\right\},
\eea
where $u=e^{-\boldsymbol{p}^2a/2Mm}$. This must be
summed over $l$ and $k_1$ and $k_2$. 
For brevity in the discussion of these sums, denote by $A(l,M)$ the
contents of the curly braces in the last equation.  
Now consider the one loop correction to the gluon propagator,
propagating $K=T/a$ time steps. The loop starts at time $k_1a$ and ends at
$k_2a$ and is positioned at $p^+=ml$. Before introducing
the counter-term we
have the following expression for the propagator correction:
\bea
D_1(\boldsymbol{p},M,K) &=& \sum_{k_1=1}^{K-3}
\sum_{k_2=k_1+2}^{K-1} u^{k_1} 
u^{K-k_2} \frac{g^2}{8 \pi^2} \frac{u^{k_2-k_1}}{(k_2-k_1)^2} 
\sum_{l=1}^{M-1} A(l,M)\nonumber
\\
&=& \frac{g^2}{8 \pi^2} \sum_{k_1=1}^{K-3} \sum_{k_2=k_1+2}^{K-1}
u^{K} \frac{1}{(k_2-k_1)^2} \sum_{l=1}^{M-1} A(l,M)\nonumber \\
&=& u^K \frac{g^2}{8\pi^2}
\left[ \left(\frac{\pi^2}{6}-1\right)K-\ln K + {\cal O}(K^0) \right]
\sum_{l=1}^{M-1} A(l,M).
\eea
The term linear in $K$ comes from terms where the loop
is short ($k_2-k_1\ll K$) and the sum is over the possible locations
of it. It is clear that when $n$ short loops are summed over their
locations we get factors proportional to $C^nK^n/n!$ where
$C$ is the coefficient of $K$ in the above linear term. The short loop behavior
therefore exponentiates and causes a shift of the ``energy'', 
$a\boldsymbol{p}^2/2M m$, in the exponent of the free propagator. 
This shift causes a gap in the gluon energy spectrum that
is forbidden in perturbation theory by Lorentz invariance. We must
therefore attempt to cancel this linear term in $K$ order by order 
in perturbation theory with a suitable choice of counter-term.
One simple choice is a two time step short
loop of exactly the structure that went into the ``bare''
self-energy.
Then at one loop order it will be proportional to the
$k=2$ term and will have the form:
\bea
e^{-ka\boldsymbol{p}^2/2mM}\frac{g^2}{4\pi^2}\frac{\xi k}{4}
\sum_{l=1}^{M-1} A(l,M),
\eea
where we adjust $\xi$ to cancel the term proportional to $k$
in the propagator correction. 
Choosing $\xi=4(1-\pi^2/6)$
does the job and we are left with a logarithmic divergence which will
be absorbed in the wave function contribution to coupling renormalization. 
We have: $\ln k = \ln (1/a) + \ln (T)$, with $T=ka$,
the total evolution time. We can therefore absorb the divergence in
the wave function renormalization factor:
\bea
\label{selfenergy}
Z(M) =1-\frac{g^2}{8\pi^2}\ln (1/a)
\sum_{l=1}^{M-1}\left\{{2\over l}
+{2\over M-l}+\frac{F(l/M)}{M}
\right\},
\eea
where
\begin{eqnarray}
F(x)&=&2f_g(x)+N_ff_f(x)+N_sf_s(x)\label{effofx}\\
\nonumber\\
f_i(x) &=& \cases{x(1-x)-2 & for $i=g$ (gluons) \cr 
x(1-x) & for $i=s$ (scalars) \cr
1/2-x(1-x) & for $i=f$ (fermions).}
\end{eqnarray}
The first two terms in the $l$ sum produce a $\ln (1/m)$ divergence
and 
we notice the familiar entanglement of ultraviolet ($a\to0$)
and infrared ($m\to0$) divergences \cite{thornfreedom}. It was
explained how these divergences disentangle in \cite{gudmundssont} and
we will discuss this further in subsection \ref{discussion}.
In (\ref{effofx}) $N_f$ counts the total number of fermionic
states, so, for example, a single Dirac fermion in 4 space-time
dimensions has $N_f=4$. We see that supersymmetry, $N_f=N_b=2+N_s$,
kills the $l$ dependent term in the summand. If $N_f=8$ as well,
the wave function contribution to coupling renormalization (apart from
the entangled divergences) vanishes. 
This is the particle content
of ${\cal N}=4$ SUSY Yang-Mills theory.

\subsection{Correction to Cubic Vertex with External Gluons}
\label{sec6.2}
Now we turn to the contribution of the proper vertex
to coupling renormalization.
The proper one loop correction to the cubic vertex is represented by a
Feynman triangle graph appearing in the worldsheet as shown in
Fig.~\ref{triangle}, which establishes our kinematics.
\begin{figure}[ht]
\begin{center}
\psfrag{'p'}{$\boldsymbol{q},l$}
\psfrag{'p1'}{$\boldsymbol{p}_1,M_1$}
\psfrag{'p2'}{$\boldsymbol{p}_2,M_2$}
\psfrag{'p3'}{$\boldsymbol{p}_3,M_3$}
\psfrag{'k0'}{$k=0$}
\psfrag{'k1'}{$k=k_1$}
\psfrag{'k2'}{$k=k_2$}
\psfrag{'l'}{$l$}
\psfrag{'M3'}{$M$}
\psfrag{'M1'}{$M_1$}
\includegraphics[height=5cm]{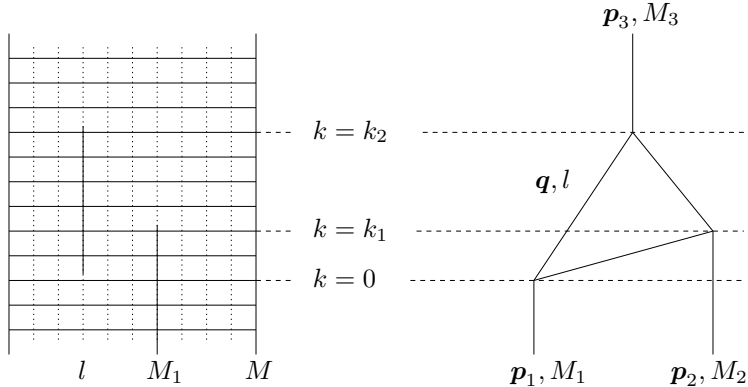}
\caption{Basic kinematic setup for the one loop correction to the
cubic vertex.
The momenta $p_1$ and $p_2$ are taken to point into the
vertex whereas $p$ points out, so that momentum conservation reads
$p_1+p_2\equiv p=-p_3$. In particular, $M\equiv M_1+M_2=-M_3$ is positive. 
We take the external gluon lines to have polarizations 
$n_1,n_2,n_3$.}
\label{triangle}
\end{center}
\end{figure}
With the external particles of Fig.~\ref{triangle} restricted
to be gluons (vector bosons) the
one loop renormalization of the gauge coupling requires calculating
the triangle graph for the different particles of the theory running
around the loop. In the following
subsections it will be useful to employ the ``complex basis''
$x^\wedge=x^1+ix^2$ and $x^\vee=x^1-ix^2$ for the first
two components of any transverse vector $\boldsymbol{x}$.

\vskip.5cm
\noindent
{\bf Fermions}
\vskip.5cm

\noindent
Referring to Appendix \ref{secb} for details of the calculation the
result for the diagram depicted in Fig.~\ref{triangle} with fermions
on the internal lines is given by:
\begin{equation}
\label{fixedlpolynomial2}
(\Gamma_1^{\rm fermions})^{\wedge \wedge \vee}
=-\frac{N_f g^3}{4 \pi^2 } \frac{a}{m} \ln
(1/a) \frac{p^+K^\wedge}{p^+_1p^+_2}
\sum_{l=1}^{M_1-1} 
\frac{f_f(l/M)}{M}
-\,(\,1 \leftrightarrow 2\,),
\end{equation}
for polarizations $n_1=n_2=\wedge,n_3=\vee$. We shall quote the 
result for other polarizations later. Recall that 
$\boldsymbol{K}$ has been defined in Eqn.~\ref{defK}. Also the
term $(1\leftrightarrow2)$ comes from the other time ordering $k_1<0$. 

\vskip.5cm
\noindent
{\bf Gluons}
\vskip.5cm

\noindent
This calculation has been done for $n_1=n_2=\wedge,n_3=\vee$ in
\cite{gudmundssont} and it is very similar to the fermion
calculation. The contribution to
charge renormalization is given by:
\bea
(\Gamma_1^{\rm gluons})^{\wedge \wedge \vee}&=&
-\frac{g^3}{4 \pi^2} \frac{a}{m} \ln(1/a) 
\frac{p^+K^\wedge}{p^+_1p^+_2}
\sum_{l=1}^{M_1-1} \Bigg\{ \frac2l
+\frac{1}{M-l}+\frac{1}{M_1-l}
+\frac{2f_g(l/M)}{M} \Bigg\}\nonumber\\
&& \qquad \qquad
-\,(\,1 \leftrightarrow 2\,).
\eea
The first three logarithmically divergent terms in the
$l$ summands again represent the entanglement of infrared and
ultraviolet divergences and we will see in section \ref{discussion}
how they cancel against 
similar terms from the self energy 
contribution. 

\vskip.5cm
\noindent
{\bf Scalars}
\vskip.5cm

\noindent
Now consider scalars on internal
lines and the same external polarizations as before. 
Recall that the indices $n_i$ in Eq.~(\ref{cartcubic}) run
from 1 to $D-2$. Let us use indices $a,b$ for directions 3 to
$D-2$. Then dimensional reduction is implemented
by taking $p_i^a=0$ for all $i$ and $a$. Using these conventions we
will be interested in the special case of Eq.~(\ref{cartcubic}) with
$n_1=a,n_2=b$ and $n_3=\vee$:
\begin{equation}
\Gamma^{ab\vee}_0=\frac{g}{8 \pi^{3/2}} \frac{a}{m} \delta^{ab}
\frac{2K^\vee}{p^+_1+p^+_2},
\end{equation}
and similarly for $\Gamma^{ab\wedge}_0$. The evaluation of the diagram
is analogous to the previous calculations and the result corresponding
to (\ref{fixedlpolynomial2}) is:
\begin{equation}
\label{fixedlpolynomial3}
(\Gamma_1^{\rm scalars})^{\wedge \wedge \vee}=
-\frac{N_s g^3}{4\pi^2}\frac{a}{m} \ln(1/a) \frac{p^+K^\wedge}{p^+_1p^+_2}
\sum_{l=1}^{M_1-1}
\frac{f_s(l/M)}{M}
-\,(\,1 \leftrightarrow 2\,).
\end{equation}

\subsection{Discussion of results}
\label{discussion}

The physical coupling can be measured by
$\prod_i \sqrt{Z_i}\Gamma$, the 
renormalized vertex function, 
where $\Gamma$ is the proper vertex and $Z_i$
is the wave function renormalization for leg $i$. To one loop we write
this in terms of 
our quantities as:
\bea
Y \equiv \Gamma_1 + \frac{1}{2} \Gamma_0 \sum_i (Z_i-1),
\eea
where $\Gamma_0$ is the tree level vertex and $\Gamma_1$ is our one
loop result for the vertex:
\bea
\Gamma_1^{\wedge \wedge \vee} &=& \left( \Gamma_1^{\rm fermions}+\Gamma_1^{\rm gluons}+\Gamma_1^{\rm
scalars}\right)^{\wedge \wedge \vee} \\
&=& -\frac{g^3}{4\pi^2}\frac{a}{m} \ln(1/a) \frac{p^+K^\wedge}{p^+_1
p^+_2}
\sum_{l=1}^{M_1-1}\Bigg(\frac{2}{l}+\frac{1}{M_1-l}+\frac{1}{M-l} +
\frac{F(l/M)}{M}\Bigg) - \,( 1 \leftrightarrow 2 ).
\nonumber
\eea
Because of how loops are treated in the BT-worldsheet
formalism it is natural to 
combine the one loop vertex result and the wave function
renormalization for each fixed position of the solid line representing
the loop. In other words we renormalize
{\em locally} on the worldsheet. To clarify this, note the
three different ways to insert a one loop correction to the cubic
vertex at fixed $l$ on the worldsheet, as in Fig.~\ref{wsfig}.
\begin{figure}[!ht]
\begin{center}
\psfrag{'M'}{$M$}
\psfrag{'M1'}{$M_1$}
\psfrag{'l'}{$l$}
\psfrag{'k0'}{$k=0$}
\psfrag{'k1'}{$k=k_1$}
\psfrag{'k2'}{$k=k_2$}
${\includegraphics[width=4cm]{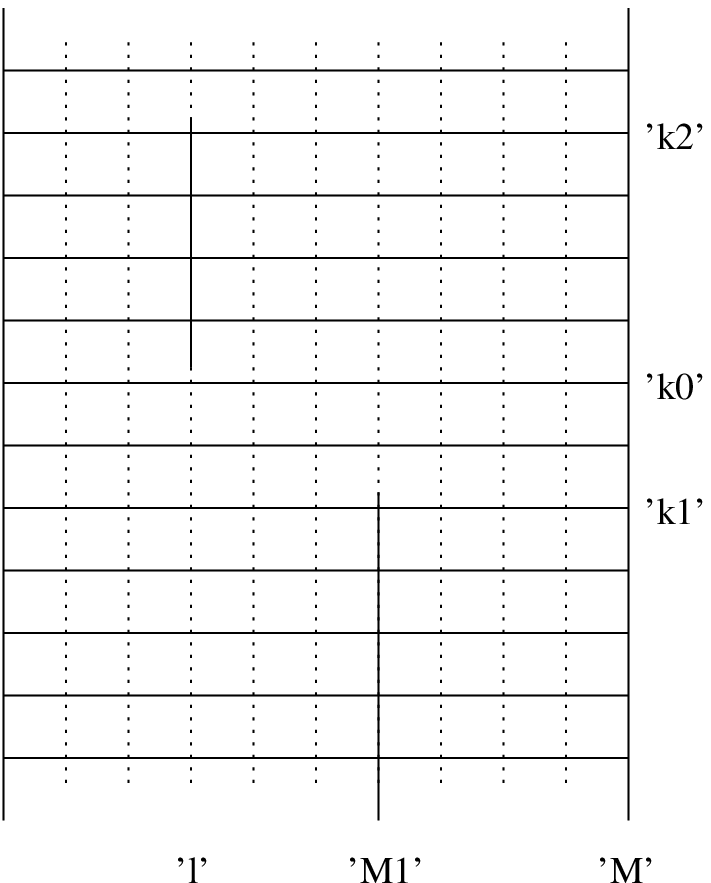}}
\hskip1cm\quad
{\includegraphics[width=4cm]{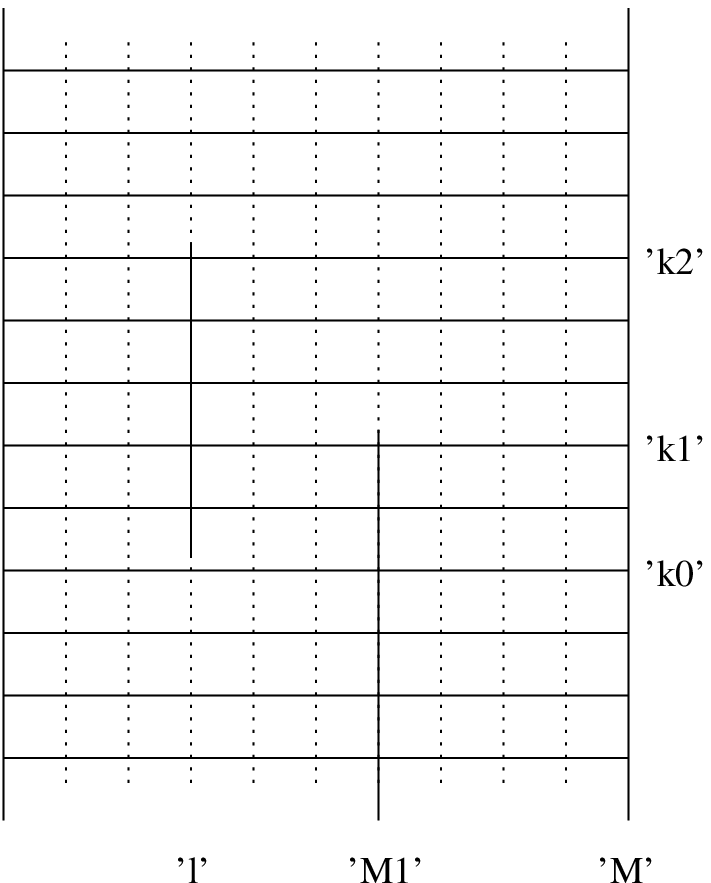}}
\hskip1cm\quad
{\includegraphics[width=4cm]{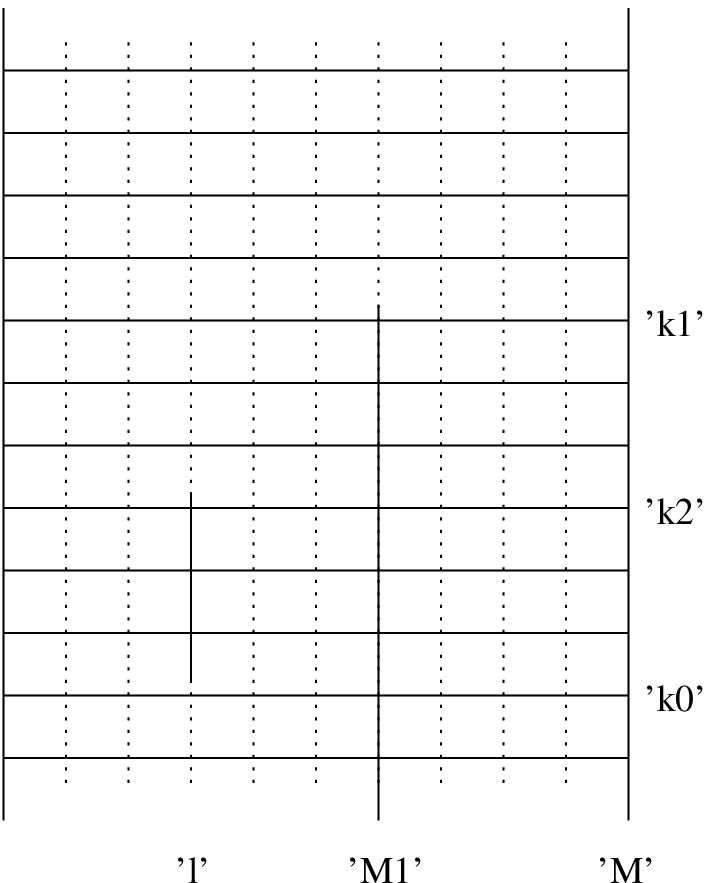}}
$
\end{center}
\caption{One loop diagrams for fixed $l$ in the BT worldsheet
picture.}
\label{wsfig}
\end{figure}
Notice that the first and last figures correspond
to self energy diagrams for 
the legs with momenta $(\boldsymbol{p},M)$ and $(\boldsymbol{p}_1,M_1)$
respectively. However, the middle figure 
corresponds to a triangle diagram with time ordering
$k_1>0$. So combining our previous results we obtain for the $Y$'s
corresponding to this polarization for each fixed $l<M_1$:
\bea
Y^{\wedge \wedge \vee}_l &=& 
\Gamma^{\wedge \wedge \vee}_{1,l} +\frac12 
\Big( 
-2g \frac{a}{m} \frac{p^+K^\wedge}{p^+_1p^+_2}
\Big)\Big(Z_l(M)-1+Z_l(M_1)-1\Big)\nonumber \\
&=&
-\frac{g^3}{4\pi^2}\frac{a}{m} \ln(1/a) \frac{p^+K^\wedge}{p^+_1
p^+_2}
\Bigg(\frac{2}{l}+\frac{1}{M_1-l}+\frac{1}{M-l} +
\frac{F(l/M)}{M}\Bigg)\nonumber \\
&& + \frac12 \Big( 
-2g \frac{a}{m} \frac{p^+K^\wedge}{p^+_1p^+_2}
\Big) \Bigg(
-\frac{g^2}{8\pi^2} \ln(1/a) \Bigg\{
\frac2l+\frac2{M-l}+\frac{F(l/M)}{M} \Bigg\} \nonumber \\
&& \phantom{+ \frac12 \Big( 
-2g \frac{a}{m} \frac{p^+K^\wedge}{p^+_1p^+_2}
\Big) \Bigg(}
-\frac{g^2}{8\pi^2} \ln(1/a) \Bigg\{
\frac2l+\frac2{M_1-l}+\frac{F(l/M_1)}{M_1} \Bigg\}
\Bigg) \nonumber \\
&=& -\frac{g^3}{8 \pi^2}
\frac{a}{m}\ln(1/a) \frac{p^+K^\wedge}{p^+_1p^+_2} 
\Bigg(
\frac{F(l/M)}{M}-\frac{F(l/M_1)}{M_1}\Bigg). 
\eea
We see that the terms of the
form $1/l, 1/(M_1-l)$ and $1/(M-l)$ cancel in the final expression
for $Y$. 
These terms multiply the $\ln(1/a)$ factor and would
result in $\ln(1/m)$ factors if the sum is taken before $\Gamma$ and
$\sqrt{Z}$ are combined. 
They represent the entanglement of $m\rightarrow0$
with $a\rightarrow0$ divergences and we have seen how this
entanglement of divergences disappears locally on the worldsheet.

Finally, for completeness we present the 
analogous results for the triangle diagram with
general polarization. The entangled divergence 
does not depend on the
polarization of the external gluons. The local disentanglement discussed
above therefore goes through unchanged for all polarizations. 
We write out the results for the renormalized vertex $Y$
where a subscript $\pm$
refers to the two different time orderings, $k_1>0, (l<M_1)$ or
$k_1<0, (l>M_1)$ respectively. 
\begin{eqnarray}
Y_{\pm}^{\wedge \wedge \wedge} &=& Y_{\pm}^{\vee \vee
\vee}=0,
\\
Y_{+}^{\wedge \wedge \vee}&=&
-\frac{g^3}{8 \pi^2}
\frac{a}{m}\ln(1/a) \frac{p^+K^\wedge}{p^+_1p^+_2} 
\sum_{l=1}^{M_1-1}\Bigg( \frac{F(l/M)}{M}-\frac{F(l/M_1)}{M_1}\Bigg),\\
Y_{+}^{\wedge \vee \wedge}&=&
-\frac{g^3}{8 \pi^2}
\frac{a}{m}\ln(1/a) \frac{p^+_2K^\wedge}{p^+_1p^+} 
\sum_{l=1}^{M_1-1}
\Bigg( -\frac{F(l/M)}{M}-\frac{F(l/M_1)}{M_1}\Bigg),
\label{udup} \\
Y_{+}^{\vee \wedge \wedge}&=&
-\frac{g^3}{8 \pi^2}
\frac{a}{m}\ln(1/a) \frac{p^+_1K^\wedge}{p^+_2p^+} 
\sum_{l=1}^{M_1-1}\Bigg(
\frac{F(l/M_1)}{M_1}-\frac{F(l/M)}{M}\Bigg),
\label{duup} \\
Y_{-}^{\wedge \wedge \vee}&=&
-\frac{g^3}{8 \pi^2}
\frac{a}{m}\ln(1/a) \frac{p^+K^\wedge}{p^+_1p^+_2} 
\sum_{l=M_1+1}^{M-1}\Bigg(
\frac{F(l/M)}{M}-\frac{F((l-M_1)/M_2)}{M_2}\Bigg),\\
Y_{-}^{\wedge \vee \wedge}&=&
-\frac{g^3}{8 \pi^2}
\frac{a}{m}\ln(1/a) \frac{p^+_2K^\wedge}{p^+_1p^+} 
\sum_{l=M_1+1}^{M-1}\Bigg(
\frac{F((l-M_1)/M_2)}{M_2} - \frac{F(l/M)}{M} \Bigg),
\label{udum} \\
Y_{-}^{\vee \wedge \wedge}&=&
-\frac{g^3}{8 \pi^2}
\frac{a}{m}\ln(1/a) \frac{p^+_1K^\wedge}{p^+_2p^+} 
\sum_{l=M_1+1}^{M-1}
\Bigg(
-\frac{F(l/M)}{M}-\frac{F((l-M_1)/M_2)}{M_2}\Bigg).
\label{duum}
\end{eqnarray}
The expressions for the $Y$'s
with $\wedge, \vee \rightarrow \vee, \wedge$ are obtained
by the replacement $K^\wedge \rightarrow K^\vee$. 
We stress that the summands in the above
expressions for $Y$ are exactly the contributions of the three diagrams in
Fig.~\ref{wsfig} with the loop fixed at $l$.
 
Define the coupling renormalization $\Delta(N_f,N_f)$ by:
\bea
Y^{n_1n_2n_3} = Y_+^{n_1n_2n_3} + Y_-^{n_1n_2n_3} 
=\Gamma_0^{n_1n_2n_3} \Delta (N_f,N_s).
\eea
We then have in the limit $M_i \rightarrow +\infty$:
\bea
\Delta (N_f,N_s) = \frac{g^2}{8\pi^2}\ln(1/a) \Bigg(
\frac{11}{3}-\frac{N_f}{3} 
-\frac{N_s}{6} \Bigg),
\eea
which is the well known result. In particular we have asymptotic
freedom when $\Delta>0$, and $\Delta$ vanishes for the particle
content of 
${\cal N}=4$ supersymmetric
Yang-Mills theory, $N_f=8$ and $N_s=6$.

For some cases such as the supersymmetric ($N_f=2+N_s$) or
pure Yang-Mills ($N_f=N_s=0$)
the summands in the expressions for the
$Y$'s do not change sign over their respective ranges. 
When $N_f<8$, so that these
cases are asymptotically free, the summands on the right sides of
(\ref{duup}) and (\ref{udum})  have
a sign which works against asymptotic freedom. Since the full sum
exhibits asymptotic freedom for each polarization,
this means that the complementary time
orderings, (\ref{udup}) and (\ref{duum}), must contribute more
than their share to asymptotic freedom. 
This fact may be useful for approximations
involving selective graph summation.

\section{Discussion and Concluding Remarks}
In this article we have completed the ``constructive'' part of
the Bardakci-Thorn program to cast the Feynman
diagrams of quantum field theory in the language of
string theory. That is, we have successfully extended the
formalism to cover the full range of interesting
supersymmetric gauge theories. 

By construction our
worldsheet systems exactly reproduce planar light-cone
diagrams modulo issues associated with renormalization
and the associated counter-terms necessary to cancel
violations of Lorentz invariance that arise because
the divergences of quantum field theory can amplify
regulator artifacts, and we are working in a non-covariant
gauge. In Section 6 we analyzed renormalization to one-loop
order and confirmed that the necessary counter-term 
can be specified {\it locally} on the worldsheet. After
that, the correct renormalization of coupling was obtained.
Further, it was found that the renormalization behavior
has an interesting local interpretation on the worldsheet.
From a purely formal point of view the major issue
left unresolved in this article is how Lorentz
invariance and the renormalization
program works on the worldsheet at 2 and higher loops.
These questions are currently under active investigation.

An exciting aspect of the BT program is 
its potential application to the confinement problem of QCD.
One of the biggest challenges in this regard is to have
a model of confinement that simultaneously and systematically
incorporates the perturbative short distance properties of
QCD. This goal is the principal motivation
for gluon chain model of confinement proposed
in \cite{greensitet}. The BT worldsheet is an ideal setting
for the construction of such models since its foundations
rest explicitly on summing Feynman diagrams. The mean
field method developed in \cite{bardakcitmean} is a first step
toward understanding the nonperturbative physics inherent
in the BT worldsheet.

But the results of the present article set the stage more for a better
understanding of Maldacena duality. Indeed, since our
worldsheet construction succeeds for ${\cal N}=4$
supersymmetric Yang-Mills theory, we now have a stringy
description of the {\it weak} 't Hooft coupling limit
to complement the strong coupling description of 
IIB string theory on an AdS$_5\times$S$_5$ manifold.
So we can approach the weak coupling/strong coupling 
duality as a relation between two stringy descriptions
rather than one between stringy and field theory descriptions.
Perhaps it will be easier to probe the interpolation between
weak and strong coupling in Maldacena duality since
one can now seek a relation of apples to apples rather than 
apples to oranges. Features of perturbative QCD such
as the scaling behavior of deep inelastic structure
functions (or the anomalous dimensions of composite
operators) can now be directly translated into the
worldsheet formalism and then given a stringy interpretation.
It will be interesting to compare this interpretation
to that given by Gubser, Klebanov, and Polyakov
in \cite{gubserkpanomalous}. 

\vskip.5cm
\noindent\underline{Acknowledgements}. We would like to thank
Zongan Qiu for helpful discussions in the early stages of this
research. This work was supported in part by the Department
of Energy under Grant No. DE-FG02-97ER-41029.

\appendix
\section{Dirac Matrices}
\label{seca}
The Dirac matrices $\Gamma^\mu$ for the $D$ dimensional Lorentz group
are $2^{D/2}\times2^{D/2}$ matrices with $\Gamma^0$ hermitian and
$\Gamma^i$ anti-hermitian, satisfying the Clifford algebra
$\{\Gamma^\mu,\Gamma^\nu\}=-2\eta^{\mu\nu}$ with 
$\eta^{\mu\nu}={\rm diag}\{-1,1,\ldots,1\}$. A spinor $\psi$
transforms under the Lorentz group by $\delta\psi=-i\epsilon\Sigma^{\mu\nu}
\psi/2$, and the conjugate spinor ${\bar \psi}$ by
$\delta{\bar \psi}=+i\epsilon{\bar \psi}\Sigma^{\mu\nu}/2$. Here
$\Sigma^{\mu\nu}={i\over 2}[\Gamma^\mu,\Gamma^\nu]$.

We choose a representation for the $\Gamma$ 
matrices that is particularly convenient for light-cone coordinates
(see \cite{bergmantbits}).
Picking the light-like directions $x^\pm=(x^0\pm x^{D-1})/\sqrt2$,
we fix $\Gamma^0$ and $\Gamma^{D-1}$ to be
\bea
\Gamma^0=i\pmatrix{0&-I\cr I&0\cr} , \qquad\qquad
\Gamma^{D-1}=i\pmatrix{0&0&{\bf 1}&0\cr0&0&0&-{\bf 1}\cr
{\bf 1}&0&0&0\cr0&-{\bf 1}&0&0\cr}\  ,
\label{gammas}
\eea
where $I$ is the $2^{(D-2)/2}$ dimensional identity matrix,
and ${\bf 1}$ is the $2^{(D-4)/2}$ dimensional identity matrix.
This will simplify the super-algebra in light-cone coordinates, 
singled out by the spatial
component $D-1$, since $\alpha^{(D-1)}$ is diagonal:
\bea
\alpha^{(D-1)}\equiv\Gamma^0\Gamma^{D-1}
=\pmatrix{{\bf 1}&0&0&0\cr0&-{\bf 1}&0&0\cr0&0&-{\bf 1}&0\cr
0&0&0&{\bf 1}\cr}\  .\label{alpha}
\eea
The choice of representation for the transverse $\Gamma^k$,
$k=1,\cdots,D-2$ can vary from one dimension to another depending
on whether or not one applies Majorana or Weyl constraints (or both).
We first separate the spinor components into
two groups denoted by checked and unchecked lower case Latin spinor indices, 
according to the eigenvalues of the matrix $\alpha^{D-1}$ (\ref{alpha}), 
the chirality matrix for $SO(1,1)$:
\bea
\alpha^{D-1}_{\check a\check b}&=& - \delta_{\check a\check b}\\
\alpha^{D-1}_{ab}&=& \delta_{ab}\\
\alpha^{D-1}_{a\check b}&=&\alpha^{D-1}_{\check ab}=0\  .
\eea
The checked and unchecked indices each range over
$2^{(D-2)/2}$ values (16 for $D=10$, 4 for $D=6$, and 2 for $D=4$). 
Because the transverse $\Gamma^k$ commute with $\alpha^{D-1}$, it
follows that $\Gamma^k_{\vphantom{\check a}a{\check b}}
=\Gamma^k_{\check a b}=0$. On the light-cone, the checked components of
the spinor fields are eliminated, leaving $2^{(D-2)/2}$
dimensional spinors $\psi^a$ acted on by $O(D-2)$ Dirac
matrices $\gamma^k_{ab}\equiv i\Gamma^k_{ab}$. by including the
extra $i$ in the definition of $\gamma^k$, we have rendered them hermitian
and their Clifford algebra is $\{\gamma^k,\gamma^l\}=2\delta_{kl}$.

To realize supersymmetric gauge theories in various
dimensions we usually have to restrict the spinors to be Majorana ($D=4$),
Weyl (D=4,6), or Majorana-Weyl (D=10), 
so that the number of fermions equals the
number of gauge bosons. Because we have chosen $\Gamma^0$
and $\Gamma^{D-1}$ pure imaginary, and because our light-cone
reduction picks the $+1$ eigenspace of $\Gamma^0\Gamma^{D-1}$,
these restrictions translate directly to corresponding 
restrictions on the $\gamma_k$. The Majorana representation,
possible for $D=2,4$ (mod 8) specifies the $\gamma_k$
to be real (and therefore symmetric). A Weyl friendly
representation possible for $D$ even is 
one for which the chirality matrix $\Gamma_{D+1}$ is
diagonal, and hence the same for the $O(D-2)$ chirality
$\gamma_{D-1}$:
\bea
\gamma_{D-1}=\pmatrix{{\bf 1}&0\cr0&-{\bf 1}\cr}.
\label{d-2chirality}
\eea
Imposing the Weyl constraint by
fixing the chirality of a spinor to be $\pm1$ means
keeping only the first (last) $2^{(D-4)/2}$ components.
Only if $D=2$(mod 8) is the Majorana condition
possible within the Weyl-friendly
representation just described. The Majorana representation
is also possible for $D=4$(mod 8), but then 
$\gamma_{D-1}$ won't be diagonal.
For example, in the case $D=4$, a Majorana
representation for the $O(D-2)$ gamma matrices can be taken to
be
\bea
\gamma^1&=&\sigma_1\; ,\qquad\qquad
\gamma^2=\sigma_3\; ,\qquad{\rm with}~\gamma_3=-i\gamma^1\gamma^2=-\sigma_2.
\eea 
The Weyl-friendly representation for $D=4$ would be
\bea
\gamma^1&=&\sigma_1\; ,\qquad\qquad
\gamma^2=\sigma_2\; ,\qquad{\rm with}~\gamma_3=-i\gamma^1\gamma^2=\sigma_3.
\eea 

\section{Details of One Loop Calculations}
\label{secb}

We consider the diagram depicted in Fig.~\ref{triangle} with fermions
on internal lines and gluons with polarizations $n_1,n_2$ and $n_3$ on 
external lines. 
Using the fermion vertices from section \ref{fermions}
we get:
\begin{eqnarray}
\sum_{l=1}^{M_1-1} \sum_{k_1,k_2} \int \frac{d\boldsymbol{q}}{(2 \pi)^3} \exp
\left\{ -\frac{a}{2m} \left(  
\frac{k_1(\boldsymbol{p}_1-\boldsymbol{q})^2}{M_1-l} +
\frac{k_2\boldsymbol{q}^2}{l} 
+\frac{(k_2-k_1)(\boldsymbol{p}-\boldsymbol{q})^2}{M-l}
\right) \right\} \nonumber \\
{\rm Tr} \Bigg\{
\left[ \frac{ag}{2m}(\gamma^{n_1} \gamma^r)
\left(\frac{q}{l}-\frac{p_1-q}{M_1-l}\right)^r + \frac{ag}{m}
\left(\frac{p_1-q}{M_1-l}-\frac{p_1}{M_1}  \right)^{n_1}  \right]
\phantom{\Bigg\}} \nonumber \\
\left[ \frac{ag}{2m}(\gamma^{n_3} \gamma^s)
\left(\frac{p-q}{M-l}-\frac{q}{l}\right)^s + \frac{ag}{m}
\left(\frac{q}{l}-\frac{p}{M}  \right)^{n_3}  \right]
\phantom{\Bigg\}} \nonumber \\
\label{rawfeynman}
\left[ \frac{ag}{2m}(\gamma^{n_2} \gamma^t)
\left(\frac{p_1-q}{M_1-l}-\frac{p-q}{M-l}\right)^t + \frac{ag}{m}
\left(\frac{p-q}{M-l}-\frac{p_2}{M_2}  \right)^{n_2}  \right]
\Bigg\}.
\end{eqnarray}
Notice that this is the expression associated with fermion arrows running
counterclockwise around the loop. The other diagram contributes the
same amount as this one. Also, this expression is for
$k_1>0$, the other time ordering $k_1<0$ is obtained by making the
substitution $p_1 \leftrightarrow p_2$ as in the gluon calculation of
\cite{gudmundssont}. 
We now proceed much as in that calculation by
completing the square in the exponent of Eq.~(\ref{rawfeynman}) and shifting
momentum: 
\begin{eqnarray}
&&\sum_{l=1}^{M_1-1} \sum_{k_1,k_2}\int \frac{d\boldsymbol{q}}{(2 \pi)^3}
\exp \left\{ -\frac{t_1+t_2+t_3}{2m/a}\boldsymbol{q}^2 \right\} e^{-H} 
\nonumber \\
&& \phantom{\sum_{l=1}^{M_1-1}}
{\rm Tr} \Bigg\{
\left[ -\frac{g}{2} (\gamma^{n_1} \gamma^r) \frac{\chi_1^r}{l(M_1-l)} + g
\frac{\chi_1^{n_1}}{M_1(M_1-l)} \right]
\left[ -\frac{g}{2} (\gamma^{n_3} \gamma^s) \frac{\chi_3^s}{l(M-l)} + g
\frac{\chi_3^{n_3}}{Ml} \right] \phantom{\Bigg\}}
\nonumber \\
&&
\phantom{\sum_{l=1}^{M_1-1}{\rm Tr} \Bigg\{ }
\left[ \frac{g}{2} (\gamma^{n_2} \gamma^t) \frac{\chi_2^t}{(M_1-l)(M-l)} + g
\frac{\chi_2^{n_2}}{(M-l)M_2} \right]
\Bigg\},
\end{eqnarray}
with
\begin{equation}
\chi_1^n =\frac{t_3 K^n/m}{t_1+t_2+t_3} -M_1 q^n, \quad
\chi_2^n =\frac{t_2 K^n/m}{t_1+t_2+t_3} -M_2 q^n, \quad
\chi_3^n =\frac{t_1 K^n/m}{t_1+t_2+t_3} +M q^n,
\end{equation}
\begin{equation}
t_1 = \frac{k_1}{M_1-l}, \quad 
t_2 = \frac{k_2}{l}, \quad
t_3 = \frac{k_2-k_1}{M-l},
\end{equation}
\begin{equation}
H=\frac{a}{2m}
\frac{t_1t_3\boldsymbol{p}_2^2+t_1t_2\boldsymbol{p}_1^2+t_2t_3
\boldsymbol{p}^2 }{t_1+t_2+t_3}.  
\end{equation}
In the $\boldsymbol{q}$-integral only the terms proportional to $\boldsymbol{q}^2$
times the Gaussian will exhibit $a \rightarrow 0$ divergences so we
retain only those. The general loop integral is given by:
\begin{eqnarray}
&&\int d\boldsymbol{q} \, \chi_1^i \chi_3^k \chi_2^j \exp \left\{
-\frac{t_1+t_2+t_3}{2m/a} \right\} \quad \rightarrow 
\nonumber
\\ 
&&\,\,\frac{(2m/a)^2\pi}{2(t_1+t_2+t_3)^3} \left(M_1M_2t_1(K^k/m)\delta^{ij}
-M_1Mt_2(K^j/m)\delta^{ik} - M_2Mt_3(K^i/m)\delta^{jk} \right). 
\end{eqnarray} 
(The arrow means that $a\rightarrow 0$ finite terms have been dropped.) Some
simplification can be done right away, for example the term
proportional to
$Tr(\gamma^{n_1} \gamma^{r} \gamma^{n_3} \gamma^{s} \gamma^{n_2}
\gamma^{t})$ after contracting with the momentum integral is
proportional to:
\begin{equation}
\sum_r {\rm Tr} \left(\gamma^{n_1} \gamma^{r} \gamma^{n_3} \gamma^{r}
\gamma^{n_2} (\boldsymbol{\gamma}\cdot\boldsymbol{K}) \right) = 
(4-D_0)\,{\rm Tr} \left(\gamma^{n_1} \gamma^{n_3} \gamma^{n_2} 
(\boldsymbol{\gamma}\cdot\boldsymbol{K}) \right), 
\end{equation}
where $D_0$ is the spacetime dimensionality of the
loop momentum integral, that is the reduced dimension $D_0=4$ so 
this term vanishes. Further simplifications can be seen when a
particular external polarization is chosen. 

The detailed $a \rightarrow 0$ behavior becomes apparent when 
the sum over $k_1$
and $k_2$ is done. With a little work it can be shown that:
\begin{eqnarray}
\sum_{k_1,k_2} \frac{t_1}{(t_1+t_2+t_3)^2}e^{-H} &\rightarrow&
\ln (1/a) \,\frac{l^3(M-l)(M_1-l)}{2M_1^2M}, \\
\sum_{k_1,k_2} \frac{t_2}{(t_1+t_2+t_3)^2}e^{-H} &\rightarrow&
\ln (1/a) \,\frac{l^2(M-l)(M_1-l)}{2M_1M}\left(
\frac{M_1-l}{M_1}+\frac{M-l}{M} \right), \\
\sum_{k_1,k_2} \frac{t_3}{(t_1+t_2+t_3)^2}e^{-H} &\rightarrow&
\ln (1/a) \,\frac{l^3(M-l)(M_1-l)}{2M_1M^2}. 
\end{eqnarray}
Carrying this through for the polarization $n_1=n_2=n_3=n$ yields:
\begin{equation}
\label{fixedlpolynomial1}
\frac{N_f a g^3 (K^n/m)}{32 \pi^2 m} \frac{\ln
(1/a)}{M_1M_2M}\sum_{l=1}^{M_1-1} \left\{ M \left[ 
1-2\frac{l}{M} \left( 1-\frac{l}{M}  \right)  \right] + M_1 \left[
1-2\frac{l}{M_1} \left( 1-\frac{l}{M_1}  \right)  \right] \right\}.
\end{equation}
In the continuum limit we have $\sum_{l=1}^{M_i-1} f(l/M_j)
\rightarrow M_j \int_{0}^{M_i/M_j} dx f(x)$ for any continuous
function $f$. Therefore,
after adding the
$k_1<0$ contribution and multiplying by 2 for the other orientation of
the fermion loop we obtain:
\begin{eqnarray}
\frac{N_f g^3}{16 \pi^2}K^n 
\frac{a}{m}\frac{(p^+_1)^2+(p^+_2)^2+(p^+)^2}{p^+_1p^+_2p^+}
\ln(1/a) \frac{2}{3}.
\end{eqnarray}

In contrast, the calculation and result for the
$n_1=n_2=\wedge,n_3=\vee$ polarization is a lot simpler. The
expression analogous to 
(\ref{fixedlpolynomial1}) is:
\begin{equation}
-\frac{N_f g^3}{16 \pi^2} \frac{a}{m} \ln
(1/a) \frac{p^+K^\wedge}{p^+_1p^+_2}\sum_{l=1}^{M_1-1}\frac1M \left[ 
1-2\frac{l}{M} \left( 1-\frac{l}{M}  \right)  \right].
\end{equation}
The result shown in (\ref{fixedlpolynomial2}) is obtained from this
one by adding the $k_1<0$ contribution and multiplying by two which
accounts for the other orientation of the fermion arrows in the loop.



\end{document}